\documentclass[fleqn,usenatbib]{mnras}

\usepackage[T1]{fontenc}

\newcommand{\su}{\mathrm{Mpc}/h}
\newcommand{\ku}{h/\mathrm{Mpc}}

\DeclareRobustCommand{\VAN}[3]{#2}
\let\VANthebibliography\thebibliography
\def\thebibliography{\DeclareRobustCommand{\VAN}[3]{##3}\VANthebibliography}

\usepackage{tikz}
\usetikzlibrary{shapes.geometric, arrows, positioning}

\usepackage{graphicx}	
\usepackage{amsmath}	
\usepackage{amssymb}	
\usepackage{newtxtext,newtxmath}

\usepackage[normalem]{ulem}

\tikzstyle{inp} = [trapezium, minimum width=3cm, minimum height=1cm, text centered, draw=black, fill=blue!30, align=center]
\tikzstyle{outp} = [trapezium, minimum width=3cm, minimum height=1cm, text centered, shape border rotate=180, draw=black, fill=blue!30, align=center]
\tikzstyle{process} = [rectangle, minimum width=3cm, minimum height=1cm, text centered, draw=black, fill=orange!30]
\tikzstyle{arrow} = [thick,->,>=stealth]

\newcommand{\orcid}[1]{\href{https://orcid.org/#1}{\includegraphics[width=0.7em]{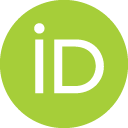}}}

\title[\textsc{FastPM} HOD modelling]{DESI Mock Challenge: Constructing DESI galaxy catalogues based on \textsc{FastPM} simulations}

\author[Andrei Variu et al.]{
\parbox{\textwidth}{
Andrei Variu\orcid{0000-0001-8615-602X},$^{1}$\thanks{E-mail: andrei.variu@epfl.ch}
Shadab Alam\orcid{0000-0002-3757-6359},$^{2,3}$\thanks{E-mail: shadab.alam@tifr.res.in}
Cheng Zhao\orcid{0000-0002-1991-7295},$^{1}$
Chia-Hsun Chuang\orcid{0000-0002-3882-078X},$^{4,5}$
Yu Yu,$^{6,7}$
Daniel Forero-Sánchez\orcid{0000-0001-5957-332X},$^{1}$
Zhejie Ding\orcid{0000-0002-3369-3718},$^{6,7}$
Jean-Paul Kneib,$^{1}$
Jessica Nicole Aguilar,$^{8}$
Steven Ahlen\orcid{0000-0001-6098-7247},$^{9}$
David Brooks,$^{10}$
Todd Claybaugh,$^{8}$
Shaun Cole\orcid{0000-0002-5954-7903},$^{11}$
Kyle Dawson,$^{4}$
Axel de la Macorra\orcid{0000-0002-1769-1640},$^{12}$
Peter Doel,$^{10}$
Jaime E. Forero-Romero\orcid{0000-0002-2890-3725},$^{13,14}$
Satya Gontcho A Gontcho\orcid{0000-0003-3142-233X},$^{8}$
Klaus Honscheid,$^{15,16,17}$
Martin Landriau\orcid{0000-0003-1838-8528},$^{8}$
Marc Manera\orcid{0000-0003-4962-8934},$^{18,19}$
Ramon	Miquel,$^{20,19}$
Jundan Nie\orcid{0000-0001-6590-8122},$^{21}$
Will Percival\orcid{0000-0002-0644-5727},$^{22,23,24}$
Claire Poppett,$^{8,25,26}$
Mehdi Rezaie\orcid{0000-0001-5589-7116},$^{27}$
Graziano Rossi,$^{28}$
Eusebio Sanchez\orcid{0000-0002-9646-8198},$^{29}$
Michael Schubnell,$^{30,31}$
Hee-Jong Seo\orcid{0000-0002-6588-3508},$^{32}$
Gregory Tarlé\orcid{0000-0003-1704-0781},$^{31}$
Mariana Vargas Magana\orcid{0000-0003-3841-1836},$^{12}$
Zhimin Zhou\orcid{0000-0002-4135-0977},$^{21}$
}
\\
\small Affiliations are listed at the end of the paper
}

\date{Accepted XXX. Received YYY; in original form ZZZ}

\pubyear{2021}

\begin{document}
\label{firstpage}
\pagerange{\pageref{firstpage}--\pageref{lastpage}}
\maketitle

\begin{abstract}
Together with larger spectroscopic surveys such as the Dark Energy Spectroscopic Instrument (DESI), the precision of large scale structure studies and thus the constraints on the cosmological parameters are rapidly improving. Therefore, one must build realistic simulations and robust covariance matrices. 
We build galaxy catalogues by applying a Halo Occupation Distribution (HOD) model upon the \textsc{FastPM} simulations, such that the resulting galaxy clustering reproduces high resolution $N$-body simulations. While the resolution and halo finder are different from the reference simulations, we reproduce the reference galaxy two-point clustering measurements -- monopole and quadrupole -- to a precision required by the DESI Year 1 Emission Line Galaxy sample down to non-linear scales, i.e. $k<0.5\,\ku$ or $s>10\,\su$.
Furthermore, we compute covariance matrices based on the resulting \textsc{FastPM} galaxy clustering -- monopole and quadrupole. We study for the first time the effect of fitting on Fourier conjugate [e.g. power spectrum] on the covariance matrix of the Fourier counterpart [e.g. correlation function]. We estimate the uncertainties of the two parameters of a simple clustering model and observe a maximum variation of 20 per cent for the different covariance matrices. Nevertheless, for most studied scales the scatter is between two to ten per cent.
Consequently, using the current pipeline we can precisely reproduce the clustering of $N$-body simulations and the resulting covariance matrices provide robust uncertainty estimations against HOD fitting scenarios. We expect our methodology will be useful for the coming DESI data analyses and their extension for other studies.
\end{abstract}




\section{Introduction}

The study of Large Scale Structure of the Universe has significantly improved in the last two decades leading to Baryon Oscillation Spectroscopic Survey \citep[BOSS;][]{2017MNRAS.470.2617A} and extended-BOSS  \citep[eBOSS;][]{2021PhRvD.103h3533A} surveys. They have published the largest 3D map of over 2 millions galaxies and quasars \citep{2021PhRvD.103h3533A}. This has allowed the measurement of cosmological parameters to a percent-level precision studying Baryonic Acoustic Oscillations (BAO) and Redshift Space Distortions (RSD). 

Currently, the Dark Energy Spectroscopic Instrument \citep[DESI;][]{2013arXiv1308.0847L,  2022AJ....164..207A} is a five years long spectroscopic survey that will outperform previous surveys by a an order of magnitude\citep{2016arXiv161100036D}, aiming to constrain the cosmological parameters with precision at a sub-percent level. With its 5000 robotically controlled optical fibres \citep{2023AJ....165....9S, 2023arXiv230606310M,2016arXiv161100037D}, DESI will scan a third of the sky to map 40 millions galaxies \citep{2023ApJ...943...68L} and quasars \citep{2023AJ....165..124A}. Only after the five-month Survey Validation \citep{2023arXiv230606307D}, DESI has measured the spectra of more than one million galaxies leading to the recent Early Data Release \citep[EDR; ][]{2023arXiv230606308D}.

Based on the DESI Legacy Imaging Surveys \citep{2017PASP..129f4101Z,2019AJ....157..168D,dr9}, there are five types of targets that are selected \citep{2023AJ....165...50M} on which optical fibres are assigned \citep{fba} to measure and analyse their spectra \citep{2023AJ....165..144G,redrock2023,2023arXiv230510426B}: Milky Way Stars \citep[MWS;][]{2020RNAAS...4..188A,2022arXiv220808514C}, Bright Galaxies \citep[BGS;][]{2020RNAAS...4..187R,2022arXiv220808512H}, Luminous Red Galaxies \citep[LRG;][]{2020RNAAS...4..181Z,2023AJ....165...58Z}, Emission Line Galaxies \citep[ELG;][]{2020RNAAS...4..180R,2023AJ....165..126R}, quasars \citep[QSO;][]{2020RNAAS...4..179Y, 2023ApJ...944..107C}. Such a complex system requires pipelines to optimise the observations \citep{2023arXiv230606309S,expcalc}.

The sub-percent precision measurements expected from ongoing and future surveys require careful analyses of the systematic effects.
To this end, the DESI Mock Challenge was launched as a series of studies and projects to build and validate the methodology for the cosmological analysis. In particular, one must find the most robust way to estimate the uncertainty of the measurements \citep{DESI_MOCK_CHALLENGE_I}.
To achieve this goal, one needs to create multiple realistic simulations of the large-scale structure, which is required to lower the noise on covariance matrix and to describe accurately the non-linear scales.

On one hand, the full $N$-body simulations  -- e.g. \citep[\textsc{SLICS};][]{2018MNRAS.481.1337H}, \citep[\textsc{UNIT};][]{Chuang:2018ega} and \citep[\textsc{AbacusSummit};][]{Maksimova:2021ynf} -- are accurate, but they are computationally expensive. Since it becomes impractical with the increase of mapped volume to have enough full $N$-body realisations to compute and test covariance matrices, they are mainly used in testing models and systematic effects. Consequently, faster but less accurate techniques have been developed -- e.g. \citep[\textsc{EZmocks};][]{Chuang:2014vfa, Zarrouk:2020hha, 2021MNRAS.503.1149Z}, \citep[\textsc{PATCHY};][]{10.1093/mnrasl/slt172},  \citep[\textsc{BAM};][]{2020MNRAS.491.2565B, 2019MNRAS.483L..58B,2020MNRAS.493..586P} -- to be run multiple times and estimate robustly the uncertainty.

In this study, we investigate the possibility to tune \textsc{FastPM} catalogues to reproduce the clustering of \textsc{SLICS} reference with the final goal of estimating the covariance matrix. In contrast to the other fast methods, \textsc{FastPM} uses accelerated particle-mesh solvers to evolve the dark-matter field, that should provide a higher accuracy of the large scale structure. The additional accuracy provided by \textsc{FastPM} can be important given the unprecedented statistical power of the DESI survey. Therefore, the final \textsc{FastPM} covariance matrix together with other methods  -- i.e. \textsc{BAM}, \textsc{EZmock}, Jackknife \citep{Zhang_jackknife_paper}, analytical models \citep{Xu13, Wad20a, Wad20b} -- are compared to the \textsc{SLICS} reference covariance matrix, in a parallel DESI Mock Challenge paper \citep{DESI_MOCK_CHALLENGE_I}.

Fundamentally similar to full $N$-body simulations, \textsc{FastPM} evolves the dark matter field into the cosmic web, the skeleton of the large scale structure in the Universe \citep[e.g.][]{2010gfe..book.....M,2018ARA&A..56..435W}. After the dark matter haloes are selected, one must implement galaxy-halo connection models \citep{2018ARA&A..56..435W} to assign galaxies.
There are more empirically inspired models such as the Halo Occupation Distribution \citep[HOD; e.g.][]{Benson2000,Seljak2000,Peacock2000,White2001,Berlind2002,Cooray2002} and Sub-Halo Abundance Matching \citep[SHAM; e.g.][]{2004ApJ...609...35K,2004ApJ...614..533T,2004MNRAS.353..189V} and more physically inspired ones such as full hydro-dynamical simulations \citep[e.g.][]{2010MNRAS.402.1536S,2015MNRAS.446..521S,dubois2014,2017MNRAS.465.2936M,2018MNRAS.473.4077P,2019MNRAS.486.2827D} or Semi Analytical Models \citep[SAMs; e.g.][]{2011MNRAS.413..101G,2014MNRAS.439..264G}. In this case, we adopt a HOD model as it is one the most efficient ways to create mock galaxy catalogues. 

Comparisons between the matter power spectra of \textsc{FastPM} and full $N$-body references have shown deviations by $\approx0.5$ per cent at $k=0.3\,\ku$ \citep{Feng2016} and by $\approx20$ per cent at $k=1\,\ku$ \citep{2022MNRAS.515.1854G}. In addition, \citet{Feng2016} have observed that the \textsc{FastPM} halo power spectrum can deviate up to $5$ per cent around $k\simeq0.5\,\ku$. Lastly, they have noticed that the \textsc{FastPM} less massive haloes are not as well matched to the reference as the higher mass haloes. As a consequence, the purpose of the current paper is to show that using a HOD model to assign galaxies on \textsc{FastPM} halo catalogues, one can overcome the differences between the \textsc{FastPM} and the full $N$-body simulations and thus reproduce the reference \textsc{SLICS} galaxy clustering. We note that the vanilla HOD model used in this work may not be sufficient to capture the physics of galaxy formation, as shown in \citep{2023arXiv230501266A, 2023MNRAS.521..937C}. Nevertheless, given the aim of this article, we are only concerned with the ability to produce the covariance matrix for two-point galaxy clustering using \textsc{FastPM}. We defer the study of other observables such as galaxy-galaxy lensing, count-in-cell statistics, and the Voronoi volume function \citep{2020MNRAS.495.3233P} to future work.

Furthermore, we compare the impact of different clustering statistics and examine the effects of various scales on the HOD fitting. Finally, we calculate covariance matrices for all the studied scenarios and perform a comparison to understand the influence of the HOD modelling on the parameter uncertainty estimation.

In Section~\ref{sec:simulations}, we present the \textsc{SLICS} and \textsc{FastPM} simulations. The methodology that we follow is detailed in Section~\ref{sec:methodology}. We describe our results on the HOD fitting performance and the covariance matrix comparison in Section~\ref{sec:results}. In the end, Section~\ref{sec:conclusion} concludes the article.

\section{Simulations}
\label{sec:simulations}

\subsection{Scinet LIght-Cone Simulations}
The Scinet LIght-Cone Simulations \citep[\textsc{SLICS};][]{2015MNRAS.450.2857H,2018MNRAS.481.1337H} consist of over 900 $N$-body mocks based on noise independent initial conditions. The large number of realisations is exploited to estimate the covariance matrices for weak lensing data \citep{2017MNRAS.465.2033J, 2017MNRAS.465.1454H, 2018MNRAS.474..712M, 2022arXiv221105779H} and for combinations of weak lensing and foreground clustering data \citep{2018MNRAS.481.5189B, 2018MNRAS.476.4662V}.

The cubic mocks -- with $L_\mathrm{box} = 505\,\su$ -- simulate a flat $\Lambda$CDM cosmology, described by the cosmology of the WMAP9 + SN + BAO, i.e. ($\Omega_{\rm m}$, $\sigma_8$, $\Omega_{\rm b}$, $w_0$, $h$, $n_{\rm s}$) = (0.2905, 0.826, 0.0447, -1.0, 0.6898, 0.969). They are obtained by running the non-linear double-mesh Poisson solver \textsc{cubep$^3$m} \citep{2013MNRAS.436..540H} to gravitationally evolve $1536^3$ particles -- with a particle mass $m_\mathrm{p} = 2.88\times 10^9~M_\odot/h$ -- on a $3072^3$ grid from $z=99.0$ up to $z=0$.

The dark matter haloes have been selected by applying a spherical over-density halo-finder \citep{2013MNRAS.436..540H}. Their mass function follows precisely the \citet{2001MNRAS.323....1S} fitting function, as shown in Figure~2 of \citet{2018MNRAS.481.1337H}. The redshift of the halo catalogues included in this study is $z=1.041$.
Lastly, given that some halo catalogues have been corrupted at the run time, we are limited to only 139 independent mocks. 

This study, together with the BAM \citep{2022arXiv221110640B} and the DESI covariance matrix comparison paper \citep{DESI_MOCK_CHALLENGE_I} are focused on the DESI Emission Line Galaxies (ELGs) sample. 
Therefore, the \textsc{SLICS} haloes have been populated using the HMQ HOD model presented in \citet{2020MNRAS.497..581A}, capable of describing the ELG galaxies -- consult \citep{2021MNRAS.504.4667A} for a comparison between different HOD models for ELGs. In addition, the parameters of the HOD model have been set to match the expected linear bias of the DESI ELG sample. The final product is a set of 139 galaxy catalogues that are used as reference in all the studies mentioned before. More details about the SLICS galaxy catalogues production can be found in the DESI covariance matrix comparison paper \citep{DESI_MOCK_CHALLENGE_I}.

\subsection{Fast Particle-Mesh}
\label{sec:fastpm}

Accelerated Particle–Mesh (PM) solvers -- such as the \textsc{FastPM} software \citep{Feng2016} -- are able to produce accurate halo populations with respect to the full $N$-body simulations. Thus, they are suitable to accurately simulate large volumes. 

\textsc{FastPM} makes use of a pencil domain-decomposition Poisson solver and Fourier-space four-point differential kernel to compute the force. Additionally, the vanilla leap-frog scheme for the time integration is adjusted to account for the acceleration of velocity during a step, allowing for the accurate tracking of the linear growth of large-scale modes regardless of the number of time steps.

For the current analysis. we have run \textsc{FastPM} with two resolutions, resulting in one set of 778 Low Resolution boxes (LR; $1296^3$ particles) and one set of 141 High Resolution (HR; $1536^3$ particles) catalogues. Both sets output snapshots at the same redshift ($z=1.041$), and have the same box side length ($L_\mathrm{box}=505\,\su$) and cosmology as the \textsc{SLICS} simulations. The particle mass of the HR simulations is $2.86444 \times 10^9~M_\odot/h$, while the one of LR is $4.77\times10^9~M_\odot/h$. The resolution of the force mesh is boosted by a factor of $B=2$ compared to the number of particles per side, for both LR and HR. Lastly, 40 linear steps have been used to evolve the density field from $a=0.05$ to $a=0.96$.

Due to the small number of \textsc{SLICS} galaxy realisations, for 123 runs of the \textsc{FastPM} (LR and HR likewise), we use the \textsc{SLICS} initial conditions. This plays an important role to reduce the effect of the cosmic variance in the clustering statistics and thus in the HOD fitting.
\textsc{SLICS} initial conditions have been built on a $1536^3$ regular grid using the Zel'dovich approximation \citep{1970A&A.....5...84Z} to displace the particles from the grid positions.
Lastly, the initial conditions have been downgraded to the LR by cutting in Fourier space the high frequency modes larger than the Nyquist frequency corresponding to the LR field.

The halos have been selected from the dark matter field with the Friends-of-Friends halo finder in \textsc{nbodykit} \citep{Hand2018}. During the galaxy assignment process -- Section~\ref{sec:hod_fitting} -- we only make use of halos with a minimum mass of $5.72 \times 10^{10}~M_\odot/h$.

Finally, in Section~\ref{sec:methodology}, when we mention \textsc{FastPM}, we imply for simplicity both HR and LR. We only make the distinction in the results section, i.e. Section~\ref{sec:results}.

\section{Methodology}
\label{sec:methodology}
\subsection{Clustering computation}

\subsubsection{Two point correlation function}
Mathematically, the two-point correlation function (2PCF) is a continuous function that can describe the clustering of galaxies. However, given the discrete nature of the galaxy distribution in the Universe, the 2PCF is measured using discrete estimators. 
In the case of cubic mocks, one can implement the natural estimator \citep{1974ApJS...28...19P}:
\begin{equation}
    \xi(s, \mu) = \frac{DD(s, \mu)}{RR(s, \mu)} - 1,
\end{equation}
where $DD(s, \mu)$ and $RR(s, \mu)$ are the data and the random pair counts, respectively, as functions of the radial distance 
\begin{equation}
s=\sqrt{s_{\perp}^2 + s_{\|}^2},
\end{equation}
and the cosine of the angle between $\bf{s}$ and the line-of-sight
\begin{equation}
    \mu=\frac{s_{\|}}{s}.
\end{equation}
In the previous equations, $s_{\perp}$ and $s_{\|}$ are the perpendicular ($\perp$) and parallel ($\|$) to the line-of-sight components of $\mathbf{s}$ , respectively. While the $DD$ term is evaluated directly on the data catalogue, $RR$ is calculated theoretically.

In the present analysis, we run \textsc{pyFCFC}\footnote{\url{https://github.com/dforero0896/pyfcfc}} the \textsc{python} wrapper of the Fast Correlation Function Calculator\footnote{\url{https://github.com/cheng-zhao/FCFC}} \citep[FCFC]{2023arXiv230112557Z} to estimate the 2PCF. Lastly, we decompose the 2D 2PCF $(\xi(s, \mu))$ into 1D multipoles $(\xi_{\ell}(s))$ with the help of the Legendre polynomials $L_{\ell}(\mu)$ of order $\ell$, as follows:
\begin{equation}
    \xi_{\ell}(s) = \frac{2\ell + 1}{2}\int_{-1}^1 \xi(s, \mu) L_{\ell}(\mu) d\mu.
\end{equation}

\subsubsection{Power spectrum}
From the mathematical point of view, the power spectrum $P(\bf{k})$ is the Fourier Transform of the 2PCF. However, the limited volume of a survey or a simulation creates mode coupling and makes the two clustering measurements not completely equivalent. Consequently, we compute the power spectrum multipoles $P_{\ell}(k)$ starting from the density field in Fourier space $\delta(\mathbf{k})$, as follows \citep{2021MNRAS.501.5616D}:
\begin{equation}
    P_{\ell}(k) = \frac{2\ell + 1}{V}\int \frac{\mathrm{d}\Omega}{4\pi} \delta(\mathbf{k})\delta(-\mathbf{k})  L_{\ell}(\hat{\mathbf{k}}\cdot\hat{\eta}) - P_{\ell}^\mathrm{noise},
\end{equation}
where the unit vector $\hat{\mathbf{k}}$ represents the direction of $\mathbf{k}$ and $\hat{\eta}$ is the global line-of-sight unit vector, chosen as the $z$ axis of the cubic simulations. Finally, given the finite galaxy number density $\bar{n}_\mathrm{g}$, the monopole shot-noise is computed as:
\begin{equation}
    P_{0}^\mathrm{noise} = \frac{1}{\bar{n}_\mathrm{g}},
\end{equation}
while for the higher-order multipoles, the shot-noise is zero.

In practice, we harness the versatility of \textsc{POWSPEC}\footnote{\url{https://github.com/cheng-zhao/powspec}} described in \citet{2021MNRAS.503.1149Z} through its \textsc{python} wrapper\footnote{\url{https://github.com/dforero0896/pypowspec}} to calculate the power spectra and their multipoles starting from the galaxy catalogues. We estimate the density field on a grid of size $512^3$, by applying the Cloud-In-Cell \citep[CIC;][]{10.1093/mnras/stw1229} particle assignment scheme on the catalogues of galaxies. Lastly, we exploit the grid interlacing technique \citep{10.1093/mnras/stw1229} to reduce the alias effects at small scales.

In the current analysis, we show the monopole ($\ell=0$), quadrupole ($\ell=2$) and hexadecapole ($\ell=4$) for both the 2PCF and the power spectrum.

\subsubsection{Bi-spectrum}
The power spectrum and the 2PCF are two-point clustering statistics, but higher order statistics are necessary to characterise more precisely the galaxy distributions. In this study, we also look at the three-point clustering statistics, namely the bi-spectrum $B(\mathbf{k_1}, \mathbf{k_2}, \mathbf{k_3})$, the Fourier pair of the three-point correlation function \citep[e.g. ][]{2002PhR...367....1B}:
\begin{equation}
\delta^D(\mathbf{k_1} + \mathbf{k_2} + \mathbf{k_3})B(\mathbf{k_1}, \mathbf{k_2}, \mathbf{k_3}) = \langle \delta(\mathbf{k_1})\delta(\mathbf{k_2})\delta(\mathbf{k_3})\rangle.
\end{equation}

The three vectors $\mathbf{k_1}, \mathbf{k_2}, \mathbf{k_3}$ are chosen to form a triangle whose two of the three sides are fixed ($k_1 = 0.1 \pm 0.05$ and $k_2 = 0.2 \pm 0.05$), but the angle $\theta_{12}$ between $\mathbf{k_1}$ and $\mathbf{k_2}$ is varied from 0 to $\pi$. In practice, we compute the monopole of the bi-spectrum on a grid of size $512^3$.

\subsection{\textsc{FastPM} HOD model}

The galaxy population and its associated clustering covariance matrix can potentially be influenced by halo properties beyond just mass, as shown in \citet{2023arXiv230501266A}. Nonetheless, such effects are expected to be small for large volume surveys such as DESI and hence we plan to address them in future work. Additionally, the \textsc{FastPM} haloes are less accurate than the ones from a $N$-body simulation, thus we do not expect that the final HOD model and parameters maintain the same physical interpretation.

As a consequence, we can adopt the simple five-parameter HOD model described in \citet{Zheng_2005} to assign galaxies to the \textsc{FastPM} halo catalogues, as long as the resulting clustering and covariance matrix match the reference. Nevertheless, in future work one can study more complex and more adapted models for the studied ELG sample.

The current model assumes that each halo can host at most one central galaxy with a probability $\mathcal{B}(1)=\langle N_\mathrm{cen} \rangle (M_{\rm h})$ dependent on the halo mass $M_{\rm h}$, where $\mathcal{B}(x)$ denotes the Bernoulli distribution and:
\begin{equation}
    \langle N_\mathrm{cen} \rangle (M_{\rm h}) = \frac{1}{2}\left[1+ \mathrm{erf}\left( \frac{\log M_{\rm h} - \log M_\mathrm{min}}{\sigma_{\log M}} \right) \right]
\end{equation}
with $\mathrm{erf}$ the error function:
\begin{equation}    
    \mathrm{erf}(x) = \frac{2}{\sqrt{\pi}} \int_0^x e^{-u^2}du.
\end{equation}
$\log M_\mathrm{min}$ is the halo mass at which the probability to host a central galaxy is one half and $\sigma_{\log M}$ controls the steepness of the transition from a probability of one to zero. Lastly, the positions and velocities of the central galaxies are precisely the values of their parent haloes.

In contrast, the number of satellite galaxies $n_\mathrm{sat}$ per halo is sampled from a Poisson distribution $\mathcal{P}(n_\mathrm{sat} | \langle N_\mathrm{sat} \rangle (M_{\rm h}))$ with the mean:
\begin{equation}
    \label{eq:hod_satelite_model}
    \langle N_\mathrm{sat} \rangle (M_{\rm h}) = \left( \frac{M_{\rm h} - M_0}{M_1} \right)^\alpha,
\end{equation}
where $M_0$ is a minimum halo mass threshold below which haloes cannot host satellite galaxies; $M_0$ and $M_1$ indicate the halo mass at which one halo hosts on average one satellite galaxy, and $\alpha$ is the power-law index. Furthermore, the positions and velocities of the satellite galaxies follow the Navarro-Frenk-White \citep[NFW]{1996ApJ...462..563N} density profile.

In the interest of adjusting the smaller scales and the quadrupole, we introduce a velocity dispersion factor ($v_\mathrm{disp}$) for the velocity parallel $(\|)$ to the line-of-sight (i.e \textbf{$\mathrm{o}Z$} in the current case) of the satellite galaxies, in addition to the five HOD parameters:
\begin{equation}
    v^\mathrm{sat, new}_{\|} =\left( v^\mathrm{sat, old}_{\|} - v^\mathrm{halo}_{\|} \right) \times v_\mathrm{disp} + v^\mathrm{halo}_{\|},
\end{equation}
where $v^\mathrm{halo}_{\|}$ is the velocity parallel to the line-of-sight of the satellites' parent halo. Finally, the six free parameters are fitted so that the resulting \textsc{FastPM} clustering matches the \textsc{SLICS} one.

\subsection{HOD fitting}
\label{sec:hod_fitting}

We would like to draw the attention of the reader to Table~\ref{tab:symbols_definitions}. It contains a summary of important symbols related to the HOD fitting.
\begin{table}
    \centering
    \begin{tabular}{c | c}
        \hline
         Notation & Meaning  \\
         \hline
         $N_\mathrm{mocks}^\mathrm{cov}=123$  & The number of \textsc{FastPM} and \textsc{SLICS} pairs \\
                                          & that share the same initial conditions.\\
                                          & These catalogues have been used \\
                                          & to compute $\textbf{C}_s$, Eq.\eqref{eq:covariance_cs}, part of $\Sigma_\mathrm{diff}$. \\
         $N_\mathrm{mocks}^\mathrm{fit}=20$  & The number of \textsc{FastPM} and \textsc{SLICS} pairs for which \\
                                        & we have computed the clustering during the  HOD  \\
                                       & fitting described in Section~\ref{sec:the_first_hod_step} and Section~\ref{sec:the_second_hod_step}.  \\
         $\Sigma_\mathrm{diag}$   &  Eq.~\eqref{eq:diagonal_cov_mat}: Diagonal matrix used during \\
                                  &  the first step of the HOD fitting, see Section~\ref{sec:the_first_hod_step}.\\
         $\sigma_{n_\mathrm{g}}$ &  Estimation of the galaxy number density noise \\
                                 & used in $\Sigma_\mathrm{diag}$. \\
                                    & Standard deviation of 139 \textsc{SLICS} mocks,\\
                                    &  divided by $\sqrt{139}$. \\ 
         $\Sigma_\mathrm{diff}$   &  Eq.~\eqref{eq:difference_cov_mat}: Difference covariance matrix used during\\
                                  &  the second step of the HOD fitting, see Section~\ref{sec:the_second_hod_step}.\\
         $\sigma'_{n_\mathrm{g}}$ &  Estimation of the galaxy number density noise\\
                                  &  used in $\Sigma_\mathrm{diff}$. \\
                                  & Standard deviation of 139 \textsc{SLICS} mocks, \\
                                  & divided by $\sqrt{N_\mathrm{mocks}^\mathrm{fit}}$. \\
          $\Sigma_\chi$           & Eq.~\eqref{eq:inverse_unbiased_covaraince}: The covariance matrix used to compute \\
                                  & the $\chi^2_\nu$, Eq.\eqref{eq:reduced_chi2}. It is not used for fitting. \\
         \hline
         
    \end{tabular}
    \caption{A summary of some of the most important and possibly confusing notations and their meaning.}
    \label{tab:symbols_definitions}
\end{table}

With the aim of finding the best-fitting \textsc{FastPM} clustering, we run a HOD Optimization Routine (\textsc{HODOR}\footnote{\url{https://github.com/Andrei-EPFL/HODOR}}). It uses the \textsc{Halotools} \citep{halotools} package to define and apply the HOD model and \textsc{PyMultiNest} \citep{2014A&A...564A.125B} the \textsc{python} wrapper of \textsc{MultiNest} \citep{2008MNRAS.384..449F,2009MNRAS.398.1601F,2019OJAp....2E..10F} to sample the six HOD parameters. 

\textsc{MultiNest} is a sampler based on Bayes' theorem that provides the maximum likelihood (best-fitting) parameters, as well as the posterior probability distribution of parameters alongside the Bayesian evidence. Bayes' theorem combines prior knowledge about the $\Theta$ parameters of a model $M$ with information from the data $D$ to calculate the posterior probability density of the $\Theta$ parameters:
\begin{equation}
    p(\Theta |D, M) = \frac{p(D | \Theta, M) p(\Theta | M)} {p(D|M)},
\end{equation}
where $p(\Theta | M)$ is the prior distribution of $\Theta$ of the model $M$, $p(D | \Theta, M)$ is the likelihood, and $p(D|M)$ is a normalizing factor called Bayesian evidence.

The uniform prior distributions that we impose on all six parameters are shown in Table~\ref{tab:parameter_priors}.
Furthermore, we approximate the likelihood by a multivariate Gaussian:
\begin{equation}
    p(D | \Theta, M) = \mathcal{L}(\Theta) \sim \mathrm{e}^{-\chi ^ 2 (\Theta) / 2},
\end{equation}
with the chi-squared:
\begin{equation}
    \chi ^ 2 (\Theta) = \mathbf{v}^\mathrm{T} \mathbf{C}^{-1} \mathbf{v},
\end{equation}
where $\mathbf{v}$ is the difference between the data and model vectors $\mathbf{v} = S_\mathrm{data} - S_\mathrm{model}(\Theta)$, and $\mathbf{C}$ is the covariance matrix.

\begin{table}
    \centering
    \begin{tabular}{c | c | c | c | c | c | c}
        \hline
         name & $\log \frac{M_\mathrm{min}}{M_\odot}$ & $\sigma_{\log M}$ & $\log \frac{M_1}{M_\odot}$ & $\kappa$ & $\alpha$ & $v_\mathrm{disp}$ \\
         \hline
         min & 11.6 & 0.01 & 9 & 0 & 0 & 0.7  \\
         max & 13.6 & 4.01 & 14 & 20 & 1.3 & 1.5  \\
         \hline
         
    \end{tabular}
    \caption{The limits of the uniform prior distributions included in the HOD fitting. Note that $M_0$ from Eq.~\eqref{eq:hod_satelite_model} is $M_0 \equiv \kappa\times M_\mathrm{min}$. $M_\odot$ denotes the solar mass.}
    \label{tab:parameter_priors}
\end{table}

The purpose of a covariance matrix $\mathbf{C}$ is to estimate the noise in the data, in the context of a noise-free model.
Nevertheless, the peculiarity of this study is that both the model ($S_\mathrm{model}(\Theta)$, \textsc{FastPM}) and the data ($S_\mathrm{data}$, \textsc{SLICS}) are affected by noise. 
Due to the small volume of the \textsc{SLICS} and \textsc{FastPM} boxes, the cosmic variance component of the noise would be larger than the expected precision of ongoing surveys such as DESI. Therefore, the simulations have been run with matching initial conditions. In this case, the relevant noise component is no longer the cosmic variance but rather the accumulated noise due to gravitational evolution while starting with exactly the same initial conditions. Thus, the mock covariance estimated by \textsc{SLICS} or \textsc{FastPM} substantially over-estimates the error for our fittings, see Figure~\ref{fig:clustering_error_bars}.

Mathematically, since one computes the difference vector $\mathbf{v}=S_\mathrm{data} - S_\mathrm{model}(\Theta)$ to estimate the $\chi^2$, one needs to estimate the covariance matrix of $\mathbf{v}$, i.e. $\mathbf{C}(\mathbf{v})$:
\begin{multline}
    \mathbf{C}(\mathbf{v}) = \mathbf{C}(S_\mathrm{data}) + \mathbf{C}(S_\mathrm{model}) \\- \mathbf{C}^\times(S_\mathrm{data}, S_\mathrm{model}) - \mathbf{C}^\times(S_\mathrm{model}, S_\mathrm{data}),
\end{multline}
where $\mathbf{C}^\times$ represents the cross covariance matrix \citep{johngubnerstats} between the data and the model vectors.
In the more common case of independent data and model vectors and noise-free model, one obtains $\mathbf{C}(\mathbf{v}) = \mathbf{C}(S_\mathrm{data})$. Nonetheless, in our case,  the \textsc{FastPM} clustering model has positively correlated noise with the \textsc{SLICS} clustering -- given the matching initial conditions -- hence $\mathbf{C}(S_\mathrm{model})\neq 0$ and $\mathbf{C}^\times \neq0$.

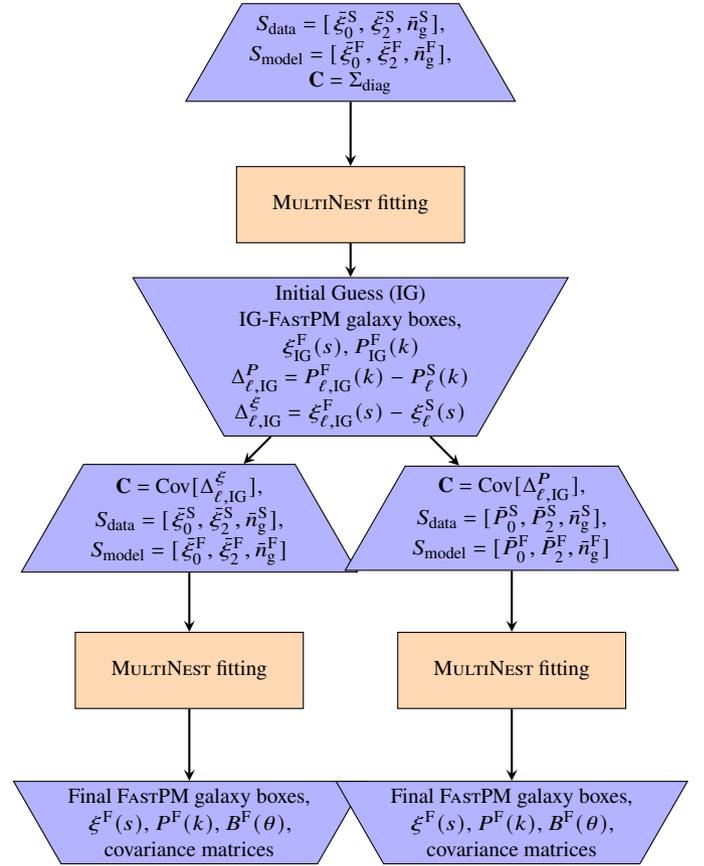
\begin{figure}
\begin{center}
\begin{tikzpicture}[auto, node distance=3cm]
\node (in1) [inp] {$S_\mathrm{data} = [\Bar{\xi}_0^\mathrm{\textsc{S}}, \Bar{\xi}_2^\mathrm{\textsc{S}}, \Bar{n}_\mathrm{g}^\mathrm{\textsc{S}}]$, \\ $S_\mathrm{model} = [\Bar{\xi}_0^\mathrm{\textsc{F}}, \Bar{\xi}_2^\mathrm{\textsc{F}}, \Bar{n}_\mathrm{g}^\mathrm{\textsc{F}}]$, \\ $\mathbf{C}=\Sigma_\mathrm{diag}$};
\node (proc1) [process, below of=in1, yshift=+1cm] {\textsc{MultiNest} fitting};
\node (ou1) [outp, below of=proc1, yshift=+1cm] {Initial Guess (IG) \\ IG-\textsc{FastPM} galaxy boxes, \\ $\xi^\textsc{F}_\mathrm{IG}(s)$, $P^\textsc{F}_\mathrm{IG}(k)$ \\ $\Delta_{\ell, \mathrm{IG}}^P = P_{\ell, \mathrm{IG}}^\mathrm{\textsc{F}}(k) - P_\ell^\mathrm{\textsc{S}}(k)$ \\ $\Delta_{\ell, \mathrm{IG}}^\xi = \xi_{\ell, \mathrm{IG}}^\mathrm{\textsc{F}}(s) - \xi_\ell^\mathrm{\textsc{S}}(s)$};
\node (in2) [inp, below left of=ou1] {$\mathbf{C}=\mathrm{Cov}[\Delta_{\ell, \mathrm{IG}}^\xi]$, \\ $S_\mathrm{data} = [\Bar{\xi}_0^\mathrm{\textsc{S}}, \Bar{\xi}_2^\mathrm{\textsc{S}}, \Bar{n}_\mathrm{g}^\mathrm{\textsc{S}}]$, \\ $S_\mathrm{model} = [\Bar{\xi}_0^\mathrm{\textsc{F}}, \Bar{\xi}_2^\mathrm{\textsc{F}}, \Bar{n}_\mathrm{g}^\mathrm{\textsc{F}}]$};
\node (in3) [inp, below right of=ou1] {$\mathbf{C}=\mathrm{Cov}[\Delta_{\ell, \mathrm{IG}}^P]$, \\ $S_\mathrm{data} = [\Bar{P}_0^\mathrm{\textsc{S}}, \Bar{P}_2^\mathrm{\textsc{S}}, \Bar{n}_\mathrm{g}^\mathrm{\textsc{S}}]$, \\ $S_\mathrm{model} = [\Bar{P}_0^\mathrm{\textsc{F}}, \Bar{P}_2^\mathrm{\textsc{F}}, \Bar{n}_\mathrm{g}^\mathrm{\textsc{F}}]$};

\node (proc2) [process, below of=in2, yshift=+1cm] {\textsc{MultiNest} fitting};
\node (ou2) [outp, below of=proc2, yshift=+1cm] {Final \textsc{FastPM} galaxy boxes, \\ $\xi^\textsc{F}(s)$, $P^\textsc{F}(k)$, $B^\textsc{F}(\theta)$, \\ covariance matrices};
\node (proc3) [process, below of=in3, yshift=+1cm] {\textsc{MultiNest} fitting};
\node (ou3) [outp, below of=proc3, yshift=+1cm] {Final \textsc{FastPM} galaxy boxes, \\ $\xi^\textsc{F}(s)$, $P^\textsc{F}(k)$, $B^\textsc{F}(\theta)$, \\ covariance matrices};

\draw [arrow] (in1) -- (proc1);
\draw [arrow] (proc1) -- (ou1);
\draw [arrow] (ou1) -- (in2);
\draw [arrow] (ou1) -- (in3);
\draw [arrow] (in2) -- (proc2);
\draw [arrow] (in3) -- (proc3);
\draw [arrow] (proc2) -- (ou2);
\draw [arrow] (proc3) -- (ou3);

\end{tikzpicture}
\end{center}
\caption{The two-step HOD fitting process that is detailed in Section~\ref{sec:hod_fitting}.}
\label{fig:flowchart_twosteps}
\end{figure}

Consequently, in order to more appropriately estimate the noise -- i.e. an estimation of $\mathbf{C}(\mathbf{v})$ -- we perform a two-step HOD fitting as schematically shown in Figure~\ref{fig:flowchart_twosteps}:
\begin{enumerate}
    \item we fit the monopole and quadrupole of the 2PCF $[\xi_0, \xi_2]$ and the galaxy number density $n_\mathrm{g}$ using a diagonal covariance matrix $(\Sigma_\mathrm{diag})$ and thus obtain an initial-guess (IG) best-fitting \textsc{FastPM} galaxy catalogues (IG-\textsc{FastPM}), see Section~\ref{sec:the_first_hod_step}; 
    \item we compute the differences $[\Delta_0, \Delta_2]$ between the clustering (monopole, quadrupole) of the IG best-fitting \textsc{FastPM} and the \textsc{SLICS} galaxy catalogues; we use these differences to calculate a new covariance matrix $(\Sigma_\mathrm{diff})$ with which we perform again the fitting, see Section~\ref{sec:the_second_hod_step}.
\end{enumerate}
In both cases, we use 20 \textsc{FastPM} (F) and 20 \textsc{SLICS} (S) halo boxes $(N_\mathrm{mocks}^\mathrm{fit} = 20)$ -- sharing the same initial conditions -- for the purpose of additionally decreasing the noise. Nonetheless, the average $\Bar{n}_\mathrm{g}^\mathrm{S}$ is computed using 139 realisations, while the average $\Bar{n}_\mathrm{g}^\mathrm{F}$ is calculated using the 20 realisations included in the HOD fitting. There are three main reasons behind this discrepancy: first, it quickly becomes expensive to apply galaxies using HOD to more than 20 \textsc{FastPM} simulations; second, the number of \textsc{SLICS} reference simulations has to be the same as for \textsc{FastPM}, so that the cosmic variance is reduced in the clustering by the shared initial conditions;  third, the noise in the galaxy number density is not reduced by the shared initial conditions, thus one needs more realisations to estimate a (practically) noiseless \textsc{SLICS} reference galaxy number density. The galaxy number density is an important constraint as it governs the shot-noise which has a significant role in the covariance matrix.

\subsubsection{The First Step} 
\label{sec:the_first_hod_step}
Initially, we perform the HOD fitting on the monopole and the quadrupole of the 2PCF, together with the galaxy number density. Hence, the data vector $S_\mathrm{data}$ is formed by concatenating their respective averages for the \textsc{SLICS} (S) mocks: $S_\mathrm{data} = [\Bar{\xi}_0^\mathrm{\textsc{S}}, \Bar{\xi}_2^\mathrm{\textsc{S}}, \Bar{n}_\mathrm{g}^\mathrm{\textsc{S}}]$. Similarly, the model vector $S_\mathrm{model}$ is determined from the \textsc{FastPM} (F) boxes: $S_\mathrm{model} = [\Bar{\xi}_0^\mathrm{\textsc{F}}, \Bar{\xi}_2^\mathrm{\textsc{F}}, \Bar{n}_\mathrm{g}^\mathrm{\textsc{F}}]$. 

Considering that the computing time of clustering measurements scales with the maximum separation, we need a large enough upper-limit to constrain relevant parameters, but small enough to keep a reasonable execution time for model evaluation during the HOD fitting. Additionally, since we are interested in capturing the non-linear effects, the lower-limit is set to 0. Consequently, the monopole and the quadrupole of the 2PCF are evaluated for $s \in [0, 50]\,\su$, with a bin size of $5\,\su$. Thus, $s$ is an array containing 10 elements $(s_1, \ldots, s_{10})$. 

As previously argued, in the first step, there is no appropriate noise estimation. Therefore, we can use an approximate covariance matrix that enables us to proceed to the second step and calculate a more suitable one. In this regard, we create a diagonal covariance matrix:
\begin{equation}
\label{eq:diagonal_cov_mat}
\Sigma_\mathrm{diag}=
    \begin{pmatrix}

    \sigma^2_1 &        &               &            &        &               &                        \\
               & \ddots &               &            &        &               &                        \\
               &        & \sigma^2_{10} &            &        &               &                        \\
               &        &               & \sigma^2_1 &        &               &                        \\
               &        &               &            & \ddots &               &                        \\
               &        &               &            &        & \sigma^2_{10} &                        \\
               &        &               &            &        &               & \sigma^2_{n_\mathrm{g}}\\
    
    \end{pmatrix},
\end{equation}
where the first 20 elements are defined as follows:
\begin{equation}
    \sigma_i=\frac{3}{s^2_i}, ~ i = 1, \ldots, 10.
\end{equation}
This selection of the diagonal covariance matrix is based on an examination of the $s^2\sigma_\mathrm{SLICS}(s)$ values, where $\sigma_\mathrm{SLICS}(s)$ represents the standard deviation of the \textsc{SLICS} 2PCF. Notably, the highest value is approximately three; hence, we initially approximate all values as three for simplicity.

The last element $\sigma_{n_\mathrm{g}}$ is computed as the standard deviation of 139 \textsc{SLICS} galaxy number densities, divided by $\sqrt{139}$, so that it estimates the uncertainty corresponding to the average of $139$ realisations. The strong constraint on the $n_g$ improves the fitting time, as \textsc{HODOR} initially evaluates the goodness-of-fit based only on the $\Bar{n}_\mathrm{g}^\mathrm{F}$ and $\Bar{n}_\mathrm{g}^\mathrm{S}$, and does not compute the clustering if $\Bar{n}_\mathrm{g}^\mathrm{F}$ is $10\sigma$ away from the reference. Additionally, the lack of covariance terms in the covariance matrix should, as well, decrease the convergence time.

Finally, we apply the best-fitting HOD model to all $N_\mathrm{mocks}^\mathrm{cov}=123$ \textsc{FastPM} halo boxes that share the initial conditions with the \textsc{SLICS} mocks to obtain the IG-\textsc{FastPM}.

\subsubsection{The Second Step}
\label{sec:the_second_hod_step}
To examine the influence of smaller scales on the HOD fitting, we compute the following for both \textsc{SLICS} and \textsc{FastPM}:
\begin{enumerate}
    \item the power spectrum for $k \in [0.02,\,k_\mathrm{max}]\,\ku$, with a bin size of $0.02\,\ku$,
    \item the 2PCF for $s \in [s_\mathrm{min},\,50]\,\su$, with a bin size of $5\,\su$,
\end{enumerate}
where the values of $k_\mathrm{max}$ and $s_\mathrm{min}$ are presented in Table~\ref{tab:fitting_intervals_n_bins}.
Consequently, we create the data and model vectors as follows:
\begin{enumerate}
    \item $S_\mathrm{data} = [\Bar{P}_0^\mathrm{\textsc{S}}, \Bar{P}_2^\mathrm{\textsc{S}}, \Bar{n}_\mathrm{g}^\mathrm{\textsc{S}}]$ and $S_\mathrm{model}=[\Bar{P}_0^\mathrm{\textsc{F}}, \Bar{P}_2^\mathrm{\textsc{F}}, \Bar{n}_\mathrm{g}^\mathrm{\textsc{F}}]$;
    \item $S_\mathrm{data} = [\Bar{\xi}_0^\mathrm{\textsc{S}}, \Bar{\xi}_2^\mathrm{\textsc{S}}, \Bar{n}_\mathrm{g}^\mathrm{\textsc{S}}]$ and $S_\mathrm{model}=[\Bar{\xi}_0^\mathrm{\textsc{F}}, \Bar{\xi}_2^\mathrm{\textsc{F}}, \Bar{n}_\mathrm{g}^\mathrm{\textsc{F}}]$.
\end{enumerate}

\begin{table}
    \centering
    \begin{tabular}{c | c | c | c}
        \hline
        name & Large & Medium & Small \\
        \hline
         $k_\mathrm{max}\,[\ku]$ & 0.5 & 0.4 & 0.3 \\
         $N^\ell_\mathrm{bins}$ & 24 & 19 & 14\\
         \hline
         $s_\mathrm{min}\,[\su]$  & 0 & 5  & 10 \\
         $N^\ell_\mathrm{bins}$ & 10 & 9 & 8\\
         \hline
         
    \end{tabular}
    \caption{The fitting ranges for the HOD fitting process described in Section~\ref{sec:the_second_hod_step}: $k \in [0.02,\,k_\mathrm{max}]\,\ku$ and $s \in [s_\mathrm{min},\,50]\,\su$. $N^\ell_\mathrm{bins}$ is the number of bins per multipole $\ell$.}
    \label{tab:fitting_intervals_n_bins}
\end{table}

In order to estimate the noise in the context of shared initial conditions between \textsc{SLICS} and \textsc{FastPM}, we use the $N_\mathrm{mocks}^\mathrm{cov}$ galaxy boxes of both \textsc{SLICS} and IG-\textsc{FastPM}, along with their corresponding clustering measurements (power spectrum or 2PCF). Furthermore, we introduce $\Delta_{\ell, \mathrm{IG}}^P = P_{\ell, \mathrm{IG}}^\mathrm{\textsc{F}}(k) - P_\ell^\mathrm{\textsc{S}}(k)$ and $\Delta_{\ell, \mathrm{IG}}^\xi = \xi_{\ell, \mathrm{IG}}^\mathrm{\textsc{F}}(s) - \xi_\ell^\mathrm{\textsc{S}}(s)$  as well as the generic vector $\Delta^\mathrm{IG}(x) = [\Delta_{0,\mathrm{IG}}, \Delta_{2,\mathrm{IG}}]$ to express the difference between the \textsc{SLICS} and the IG-\textsc{FastPM} galaxy clustering that share the initial conditions. Here, the variable $x$ represents either $k$ or $s$.

Taking advantage of the previous definitions, we further define a matrix \textbf{M} with the following elements:
\begin{equation}
    \textbf{M}_{ij}=\Delta_i^\mathrm{IG}(x_j) - \Bar{\Delta}^\mathrm{IG}(x_j),~i = 1, 2, ..., N_\mathrm{mocks}^\mathrm{cov},~x_j\in[x_\mathrm{min}, x_\mathrm{max}],
\end{equation}
where $\Delta_i^\mathrm{IG}$ denotes the vector corresponding to the $i-$th (\textsc{SLICS}, IG-\textsc{FastPM}) pair, $\Bar{\Delta}^\mathrm{IG}$ represents the mean vector over all (\textsc{SLICS}, IG-\textsc{FastPM}) pairs and $[x_\mathrm{min}, x_\mathrm{max}]$ defines the interval of points involved in the fitting, see Table~\ref{tab:fitting_intervals_n_bins}. Starting from this matrix and its transpose, we calculate the sample covariance matrix $\textbf{C}_s$ as follows:
\begin{equation}
    \label{eq:covariance_cs}
    \textbf{C}_s = \frac{1}{N_\mathrm{mocks}^\mathrm{cov} - 1} \textbf{M}^\mathrm{T}\textbf{M}.
\end{equation}

Lastly, we calculate the $\sigma'_{n_\mathrm{g}}$ as the standard deviation of 139 \textsc{SLICS} galaxy number densities, divided by $\sqrt{N_\mathrm{mocks}^\mathrm{fit}}$ -- so that it estimates the uncertainty corresponding to the average of $N_\mathrm{mocks}^\mathrm{fit}$ realisations -- and we attach it to the $\textbf{C}_s$ to obtain the final covariance matrix used in the HOD fitting:
\begin{equation}
\label{eq:difference_cov_mat}
\Sigma_\mathrm{diff}\equiv
    \begin{pmatrix}

    \textbf{C}_s &     0                     \\
           0     & \sigma'^2_{n_\mathrm{g}}  \\
    \end{pmatrix}.
\end{equation}
Note that while the error estimate for the clustering is based on the difference in clustering due to matched initial condition, the error of the number density is directly computed from the \textsc{SLICS} realisations, as we aim to constrain the absolute number density, which has strong effect on the final clustering covariance.

\subsubsection{Goodness-of-fit}
\label{sec:reduced_chi2}
In this section, we define a reduced $\chi^2$ -- $\chi^2_\nu$ -- that expresses the goodness-of-fit for the average of $N_\mathrm{mocks}^\mathrm{fit}$ \textsc{FastPM} galaxy clustering realisations with respect the \textsc{SLICS} reference, i.e. the $n_\mathrm{g}$ is not included:
\begin{equation}
    \label{eq:reduced_chi2}
    \chi_\nu ^ 2 = N_\mathrm{mocks}^\mathrm{fit} \times \frac{\mathbf{\Delta}^\mathrm{T} \Sigma_\chi^{-1} \mathbf{\Delta}}{\nu},
\end{equation}
where $\mathbf{\Delta}$ denotes the difference between \textsc{FastPM} and \textsc{SLICS} clustering -- monopole and quadrupole -- and $\nu = N_\mathrm{bins} - N_\mathrm{params}$, with 
\begin{enumerate}
    \item $N_\mathrm{params}=6$ -- the number of free parameters;
    \item $N_\mathrm{bins} = 2 \times N^\ell_\mathrm{bins}$ -- the length of the $\Delta^\mathrm{IG}(x)$ vector, see Table~\ref{tab:fitting_intervals_n_bins}.
\end{enumerate}

The $\Sigma_\chi^{-1}$ is the unbiased estimate of the inverse covariance matrix \citep{2007A&A...464..399H}:
\begin{equation}
    \label{eq:inverse_unbiased_covaraince}
    \Sigma_\chi^{-1} = \textbf{C}_s^{-1}\frac{N_\mathrm{mocks}^\mathrm{cov} - N_\mathrm{bins} -2}{N_\mathrm{mocks}^\mathrm{cov} - 1}, 
\end{equation}
where $\textbf{C}_s$ is defined in Eq.~\eqref{eq:covariance_cs}. \citet{2016MNRAS.456L.132S, 2022MNRAS.510.3207P} have shown that this correction may not be the optimal choice for accurately determining the uncertainty of the parameters. However, since our main focus is on obtaining the best-fitting clustering and assessing its goodness-of-fit, it remains a reasonable correction.

Finally, as we fit the average of $N_\mathrm{mocks}^\mathrm{fit}$ realisations, we must scale the covariance matrix $\textbf{C}_s$ by a factor of $1 / N_\mathrm{mocks}^\mathrm{fit}$. As a consequence, the $N_\mathrm{mocks}^\mathrm{fit}$ factor appears in Eq.\eqref{eq:reduced_chi2}.

\subsection{Covariance matrix comparison}
\label{sec:cov_mat_constrain_power}

Given that the main goal is to have a robust estimation of the uncertainty on the cosmological parameters, we want to compare the constraining power of the covariance matrices. To this end, we fit the 123 individual \textsc{SLICS} clustering (monopole and quadrupole) with the following models:
\begin{equation}
    P^\ell_\mathrm{model}(k) = b_\ell \times \Bar{P}^\ell_\mathrm{123, SLICS}(k)
\end{equation}
and
\begin{equation}
    \xi^\ell_\mathrm{model}(s) = b_\ell \times \Bar{\xi}^\ell_\mathrm{123, SLICS}(s),
\end{equation}
where $\Bar{P}^\ell_\mathrm{123, SLICS}(k)$ and $\Bar{\xi}^\ell_\mathrm{123, SLICS}(s)$ are averages of the 123 realisations and $b_\ell$ denotes the two free parameters.

Moreover, the covariance matrices are computed similarly to the Eq.~\eqref{eq:inverse_unbiased_covaraince}, but using 778 LR \textsc{FastPM} realisations. The fitting is performed using \textsc{PyMultiNest}, for different fitting ranges ($k \in [0.02,\,\mathcal{K}]\,\ku$ and $s \in [\mathcal{S},\,200]\,\su$, see Table~\ref{tab:fitting_intervals_covariance_comparison}) for the purpose of comparing the effect of the covariance matrices at different scales. The largest fitting intervals are chosen so that they cover the nominal scales included in the BAO and RSD analyses, i.e. $\mathcal{K}\approx0.2\,\ku$ and $\mathcal{S}\approx 20 \,\su$ \citep[e.g. ][]{2020MNRAS.499.5527T, 2021MNRAS.501.5616D}. Finally, the shown values are the average ($b_\ell$) and standard deviation ($\sigma_{b_\ell}$) of the marginalised posterior $p(b_\ell)$  and covariance ($\mathcal{R}[b_0, b_2]$) of the posterior distribution of $b_0$ and $b_2$, $p(b_0, b_2)$. By construction, the values of $b_\ell$ should be one.

The main reason why we perform such a simplified test is to avoid the systematic errors that can arise due to the modelling. Consequently, the comparison between the quoted $\sigma_{b_\ell}$ and $\mathcal{R}[b_0, b_2]$ should be directly related to the differences in \textsc{FastPM} covariance matrices. We, nevertheless, reckon that these comparisons do not show how the errors on the parameters of a realistic BAO/RSD model would behave.

\begin{table}
    \centering
    \begin{tabular}{c | c | c | c | c}
        \hline
         $\mathcal{K}\,[\ku]$ & 0.1 & 0.15 & 0.2 & 0.25  \\
         \hline
         $\mathcal{S}\,[\su]$  & 15 & 20  & 25 & 30 \\
         \hline
         
    \end{tabular}
    \caption{The fitting ranges -- $k \in [0.02,\,\mathcal{K}]\,\ku$ and $s \in [\mathcal{S},\,200]\,\su$ used in the clustering fitting described in Section~\ref{sec:cov_mat_constrain_power}}
    \label{tab:fitting_intervals_covariance_comparison}
\end{table}

\section{Results}
\label{sec:tension_parameter}
\label{sec:results}

\begin{figure}
	\includegraphics[width=\columnwidth]{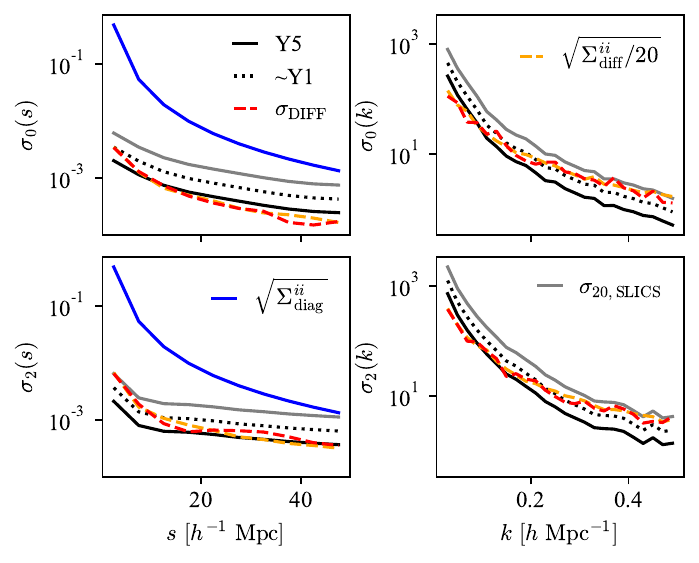}
    \caption{Monopole and quadrupole error bars: left panels -- 2PCF; right panels -- power spectrum. Black -- estimated uncertainty for the entire DESI survey (Y5); Dotted black -- estimated uncertainty for the Year 1 (Y1) DESI survey; Blue -- square root of the $\Sigma_\mathrm{diag}$'s terms; Dashed orange -- square root of the $\Sigma_\mathrm{diff}$'s diagonal terms, divided by $\sqrt{20}$, $N_\mathrm{mocks}^\mathrm{fit}=20$; Dashed red -- standard deviation of the differences between the best-fitting \textsc{FastPM} clustering (from the second HOD fitting step, one HOD fitting scenario) and \textsc{SLICS} ($N_\mathrm{mocks}^\mathrm{fit}$ realisations), further divided by $\sqrt{N_\mathrm{mocks}^\mathrm{fit}}$; Grey -- standard deviation of $N_\mathrm{mocks}^\mathrm{fit}$ \textsc{SLICS} clustering realisations, further divided by $\sqrt{N_\mathrm{mocks}^\mathrm{fit}}$. We emphasise that the Y5 and Y1 errors are estimated by scaling the \textsc{SLICS} covariance to the effective volumes of $24\,\mathrm{Gpc}^3\,h^{-3}$ and $8\,\mathrm{Gpc}^3\,h^{-3}$, respectively. }
    \label{fig:clustering_error_bars}
\end{figure}

One of the challenges of HOD fitting is addressing the high precision imposed by large volume surveys such as DESI because it requires prohibitively many large volume simulations. Figure~\ref{fig:clustering_error_bars} illustrates this issue as a comparison between $\sigma_\mathrm{20,SLICS}$\footnote{This would be the noise level in a hypothetical case where \textsc{SLICS} and \textsc{FastPM} would not share the initial conditions.} the noise corresponding to the average of $N_\mathrm{mocks}^\mathrm{fit}=20$ \textsc{SLICS} clustering realisations and the expected DESI Y5\footnote{The DESI Year 5 error is estimated by rescaling $\sigma_\mathrm{20,SLICS}$ to match the Y5 ELG sample volume, which is assumed to be $24\,\mathrm{Gpc}^3\,h^{-3}$.} and Y1\footnote{The DESI Year 1 error is estimated by rescaling $\sigma_\mathrm{20,SLICS}$ to match the Y1 ELG sample volume, which is assumed to be one third of the Y5 volume.} errors of the ELG sample. It is obvious that $N_\mathrm{mocks}^\mathrm{fit}$ \textsc{SLICS} realisations do not reach the required precision\footnote{A simple calculation reveals that one would need 192 \textsc{SLICS} realisations to meet the DESI Y5 precision requirements.}.

In order to overcome this issue, we employ the novel matched initial conditions simulations (\textsc{SLICS} and \textsc{FastPM}). In this case, the effect of the cosmic variance on the clustering difference is mostly removed. Therefore, as discussed in Section~\ref{sec:hod_fitting}, the relevant error estimate is given by the covariance matrix of the clustering difference between the two simulations. Given the fact that we use $N_\mathrm{mocks}^\mathrm{fit}$ pairs to perform the HOD fitting, the covariance matrix must be rescaled by $N_\mathrm{mocks}^\mathrm{fit}$. The square root of the diagonal of the resulting covariance matrix is illustrated with an dashed orange line in Figure~\ref{fig:clustering_error_bars}. One can observe that the matched initial conditions significantly reduce the noise to values below $\sigma_\mathrm{20,SLICS}$.

Furthermore, we would like to highlight that the precision depicted by the dashed orange line is either better than or equal to DESI Y1 precision up to $k \approx 0.25\,\ku$. Consequently, the results presented in this paper are precise enough with respect to the requirements of further DESI Y1 analyses. Nonetheless, it might be necessary to readdress this study for the full DESI sample, to account for even lower noise levels. For this, one could use the 1800 \textsc{AbacusSummit} \citep{Maksimova:2021ynf} $N$-body $0.5\,\mathrm{Gpc}/h$ cubic boxes.

In addition, Figure~\ref{fig:clustering_error_bars} illustrates the comparison between $\sigma_\mathrm{DIFF}$ and the square root of the diagonal elements of $\Sigma_\mathrm{diff}$. In this context, $\sigma_\mathrm{DIFF}$ represents the standard deviation of the differences between the best-fitting \textsc{FastPM} (obtained from the second HOD fitting step) and \textsc{SLICS} clustering, further divided by $\sqrt{N_\mathrm{mocks}^\mathrm{fit}}$. Ideally, an iterative HOD fitting process should be performed to ensure a robust $\Sigma_\mathrm{diff}$, but the close agreement between $\sigma_\mathrm{DIFF}$ and the diagonal elements of $\Sigma_\mathrm{diff}$ suggests that $\Sigma_\mathrm{diff}$ has approximately converged after a single iteration. A more detailed argument in support of the convergence of $\Sigma_\mathrm{diff}$ is presented in Section~\ref{sec:appendix_robustnesstest}.

As pointed out in Section~\ref{sec:hod_fitting}, it is important that the \textsc{FastPM} galaxy catalogues reproduce the \textsc{SLICS} shot-noise. Examining the \textsc{FastPM} galaxy number densities of all HOD fitting cases, we observed that the largest deviation, $| \Bar{n}_\mathrm{g}^\mathrm{S} - \Bar{n}_\mathrm{g}^\mathrm{F} |  / \sigma_{n_\mathrm{g}}'$, is approximately $0.5\,\sigma$, but most values are below $0.2\,\sigma$. This strongly supports that the galaxy number density is well constrained and that it is safe to define a $\chi^2_\nu$ without including $n_\mathrm{g}$ -- see Eq.\eqref{eq:reduced_chi2}.

Furthermore, the values of the $\chi^2_\nu$ are subject to uncertainties due to the finite number of realisations used to estimate the covariance matrix and the limited number of HOD realisations per halo catalogue. The most significant uncertainty, $\approx 27$ per cent, arises from the limited number of HOD realisations. The remaining values are below 20 per cent, see Section~\ref{sec:appendix_chi2uncertainty} for more details. The $\chi^2_\nu$ is simply used as a metric to evaluate the goodness-of-fit. For this reason it is important to consider that it is affected by a large uncertainty when comparing its magnitude to the expected value of one.

The primary focus of this paper is to investigate the limits of the \textsc{FastPM} capabilities to model the non-linear scales captured by $N$-body simulations. Furthermore, we study the effect of fitting to successively more non-linear scales and either Fourier or configuration space statistics on the \textsc{FastPM} covariance matrix.

\subsection{Power spectrum fitting}
\label{sec:power_spectrum_fitting}

\begin{figure*}
	\includegraphics[width=\columnwidth]{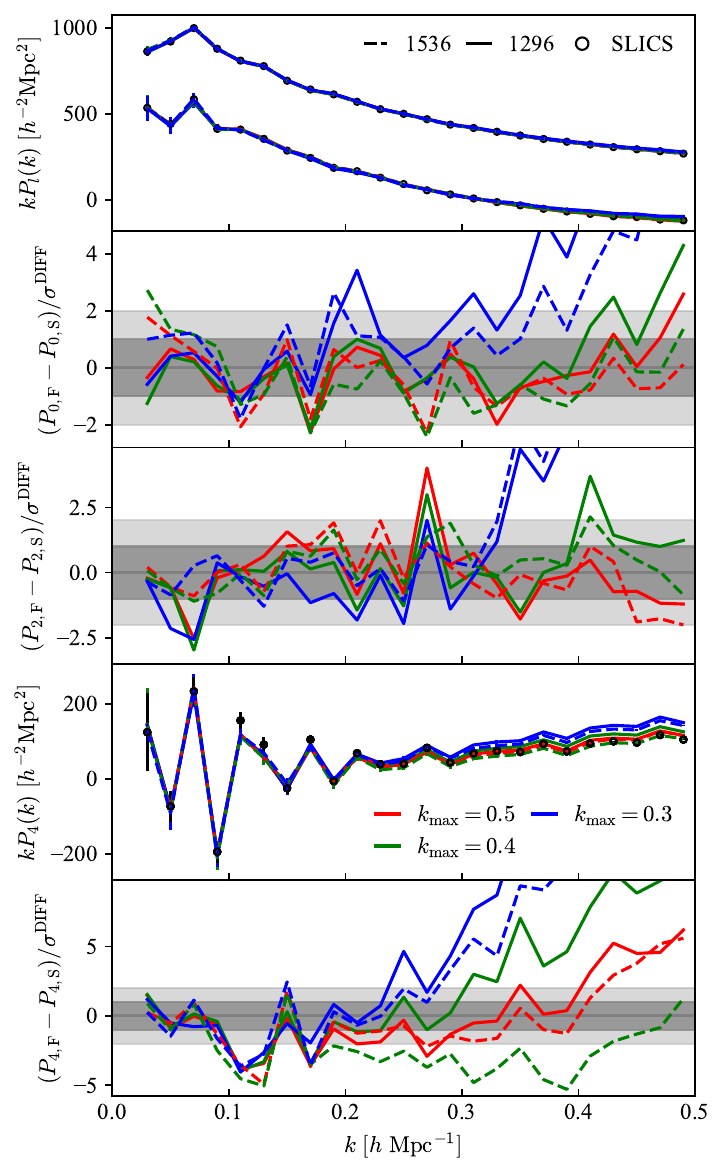}
 \includegraphics[width=\columnwidth]{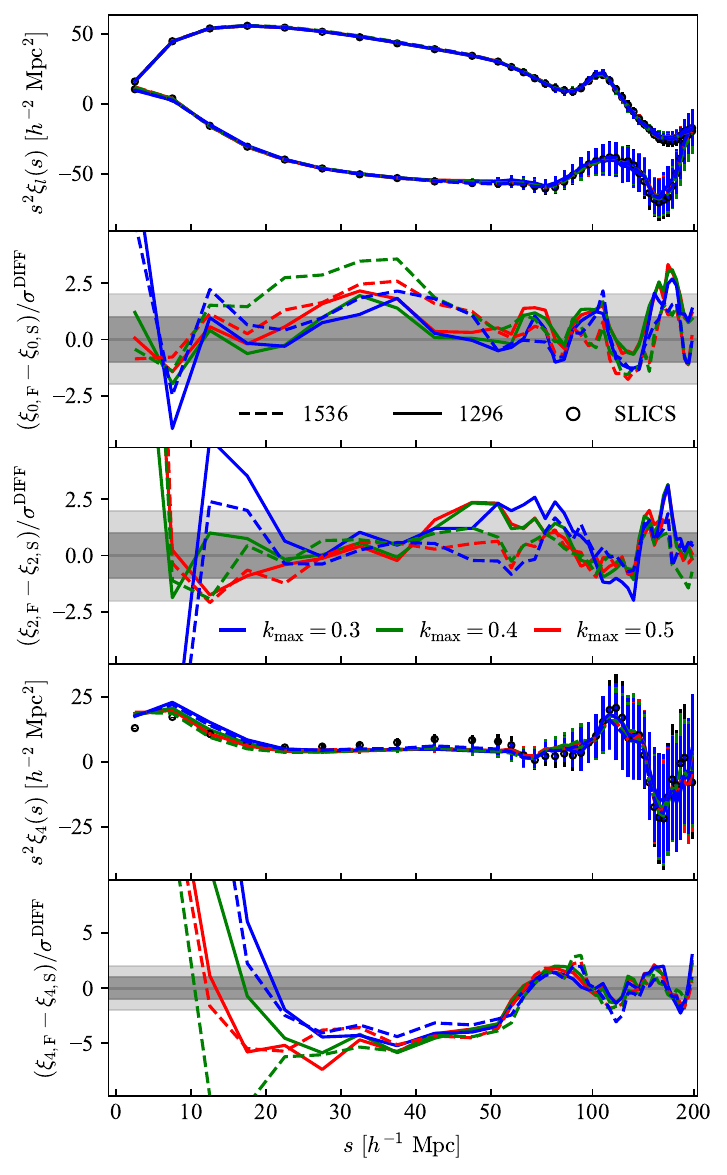}
    \caption{The average of 20 \textsc{SLICS} (reference-black) and 20 \textsc{FastPM} (model-colours) clustering realisations and the tension ($\sigma_\mathrm{DIFF}$ is shown in Figure~\ref{fig:clustering_error_bars}) between them: left - power spectrum and right - 2PCF. \textsc{FastPM} mocks share the white-noise through the initial conditions with the \textsc{SLICS} ones. The fitting has been performed: 1) on the monopole and quadrupole of the power spectrum; 2) for three different fitting ranges, see Table~\ref{tab:fitting_intervals_n_bins}; 3) using HR (dashed) and LR (continuous) \textsc{FastPM} realisations. The O$x$ axis of the 2PCF panels has a linear scale from 0 to $50\,\su$ and a logarithmic scale above this limit.}
    \label{fig:1296_1536_DIFF_PK_kmax_05_04_03}
\end{figure*}

Figure~\ref{fig:1296_1536_DIFF_PK_kmax_05_04_03} shows the results of the HOD fitting performed on the power spectrum for three different $k$ intervals, defined in Table~\ref{tab:fitting_intervals_n_bins}. The second, third and fifth rows display the difference in the clustering scaled by the difference error. We remind the reader that this error is smaller than the expected one for the given volume, due to the matched initial conditions between the two simulations, see Figure~\ref{fig:clustering_error_bars}.

The best fitting monopoles and quadrupoles are within $\pm 1 \sigma$ for most scales. Moreover, the results for the HR \textsc{FastPM} -- presented with dashed line -- are only marginally better than the ones for LR \textsc{FastPM}. Given the modest difference between the performances of the two resolutions, we believe that the LR \textsc{FastPM} is precise enough to describe the two-point clustering to non-linear scales for a DESI ELG-like galaxy sample, within the estimated DESI Y1 error bars, see Figure~\ref{fig:clustering_error_bars}.

Considering that we only fit the first two even multipoles, there is no guarantee that the third one would match the reference. Nevertheless, the fifth row of Figure~\ref{fig:1296_1536_DIFF_PK_kmax_05_04_03} illustrates that fitting the monopole and quadrupole to smaller scales improves the agreement of the hexadecapole. For instance, fitting on the Large interval pushes the $\ell=4$ multipole within $\pm2\sigma$ for $k<0.4\,\ku$, whereas for Medium and Small intervals, the hexadecapole is placed within $\pm2\sigma$ only for $k<0.3\,\ku$ or $k<0.2\,\ku$, respectively.

Due to the fact that the power spectrum is affected by the window function, it is not obvious that a good matching in Fourier space translates as a good matching in Configuration space. Thus, we compute and display the corresponding 2PCF in the right-hand side of Figure~\ref{fig:1296_1536_DIFF_PK_kmax_05_04_03}.
Most monopoles and quadrupoles agree within $\pm2\sigma$ with \textsc{SLICS} for separations larger than $20\,\su$. This suggests that it is possible to obtain a reasonable 2PCF above a certain minimum separation, even when performing HOD fitting on the power spectrum.
However, fitting on the Medium and Large intervals, the $2\sigma$ matching goes down to a separation of $10\,\su$.

In contrast, for separations smaller than $5\,\su$, the non-linear effects become dominant, making it difficult to replicate the velocity field. This is why increasing the fitting range up to $k_\mathrm{max}=0.5$ can improve the monopole but not the quadrupole.
Lastly, the 2PCF hexadecapole exhibits a bias of over $3 \sigma$ for $s<50\,\su$ in all six cases.

\begin{figure*}
	\includegraphics[width=\textwidth]{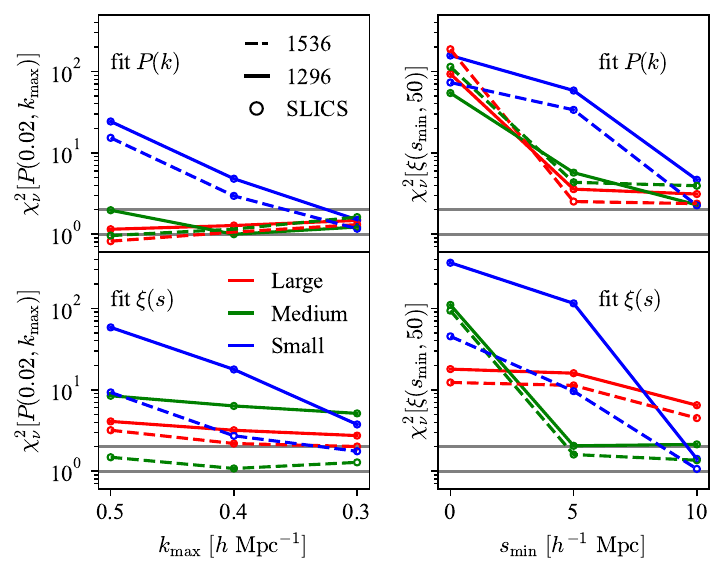}
    \caption{The $\chi_\nu^2$ as defined in Section~\ref{sec:reduced_chi2}. We compute $\chi_\nu^2$: 1) for different intervals (see $\mathrm{O}y$ and $\mathrm{O}x$ axes) of the clustering statistics (left panels - power spectrum; right panels - 2PCF); 2) for different fitted clustering (upper panels - power spectrum, see Section~\ref{sec:power_spectrum_fitting}; lower panels - 2PCF, see Section~\ref{sec:2pcf_fitting}); 3) for different fitting ranges (see Table~\ref{tab:fitting_intervals_n_bins}).}
    \label{fig:reduced_chi2}
\end{figure*}

After a more qualitative description of the results, we present the $\chi^2_\nu$ values in the upper panels of Figure~\ref{fig:reduced_chi2}. Generally, the HR \textsc{FastPM} produces lower $\chi^2_\nu$ values than the LR, as expected from Figure~\ref{fig:1296_1536_DIFF_PK_kmax_05_04_03}.
However, $\chi^2_\nu [P(0.02, k_\mathrm{max})]\simeq 1$, which reiterates that by fitting the monopole and quadrupole of the power spectrum up to the three $k_\mathrm{max}$ values, one can achieve a good match with the \textsc{SLICS} reference, within the DESI Y1 precision even with LR. In addition, $\chi^2_\nu [\xi(20, 50)]\simeq 2$ for the small fitting interval of the LR power spectrum, reinforcing the fact that one can get a reasonable 2PCF above a certain minimum separtion threshold when the fitting is performed on the power spectrum.

Additionally, we can observe the behaviour of $\chi^2_\nu$ when it is estimated on different intervals than those used for the fitting. When the fitting is performed on the Large interval, the $\chi^2_\nu\simeq 1$ for all smaller intervals, regardless of the resolution. However, fitting on the Medium interval shows that the difference between HR and LR becomes more significant for $k>0.4 \,\ku$ (see also Figure~\ref{fig:1296_1536_DIFF_PK_kmax_05_04_03}): the $\chi^2_\nu\simeq 2$ for LR, while for HR, it is close to one. These findings imply that fitting up to $k\leq0.4 \,\ku$ is satisfactory for HR \textsc{FastPM}, whereas smaller scales play a more significant role in LR.

Furthermore, fitting on the Small interval shows that although $\chi^2_\nu [P(0.02, 0.3)]\simeq 1$, it is much larger for $k>0.3 \,\ku$, indicating strong clustering divergence beyond that value (see Figure~\ref{fig:1296_1536_DIFF_PK_kmax_05_04_03}). Therefore, both LR and HR benefit from considering the clustering information contained in smaller scales $k>0.3 \,\ku$.

\subsection{2PCF fitting}
\label{sec:2pcf_fitting}

When the HOD fitting is performed on the power spectrum, the minimum 2PCF $\chi^2_\nu$ is $\chi^2_\nu[\xi(10, 50)] \approx 2$. While this translates to a $2\sigma$ agreement down to the separation of $10\,\su$ between \textsc{FastPM} and \textsc{SLICS} 2PCF, we test whether fitting directly the 2PCF can improve the results.
Therefore, in this section, we analyse the outcomes of the HOD fitting performed on the 2PCF monopole and quadrupole, for $s \in [s_\mathrm{min}, 50]\,\su$, see Table~\ref{tab:fitting_intervals_n_bins}.

\begin{figure*}
	\includegraphics[width=\columnwidth]{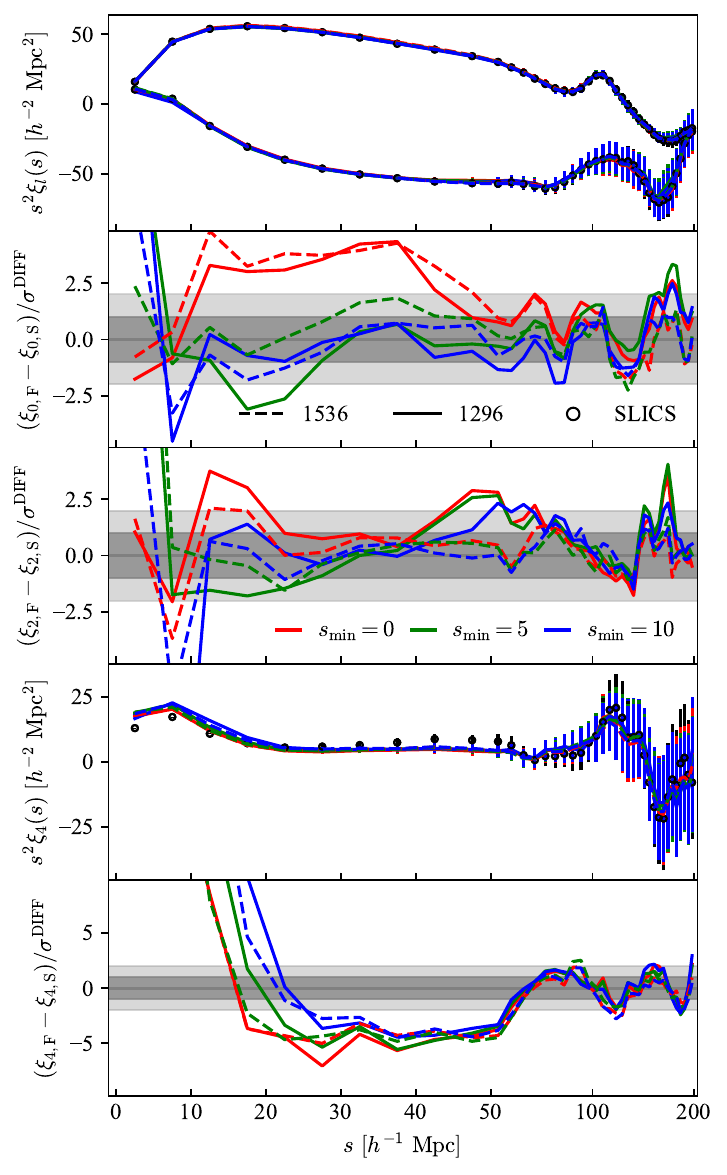}
 \includegraphics[width=\columnwidth]{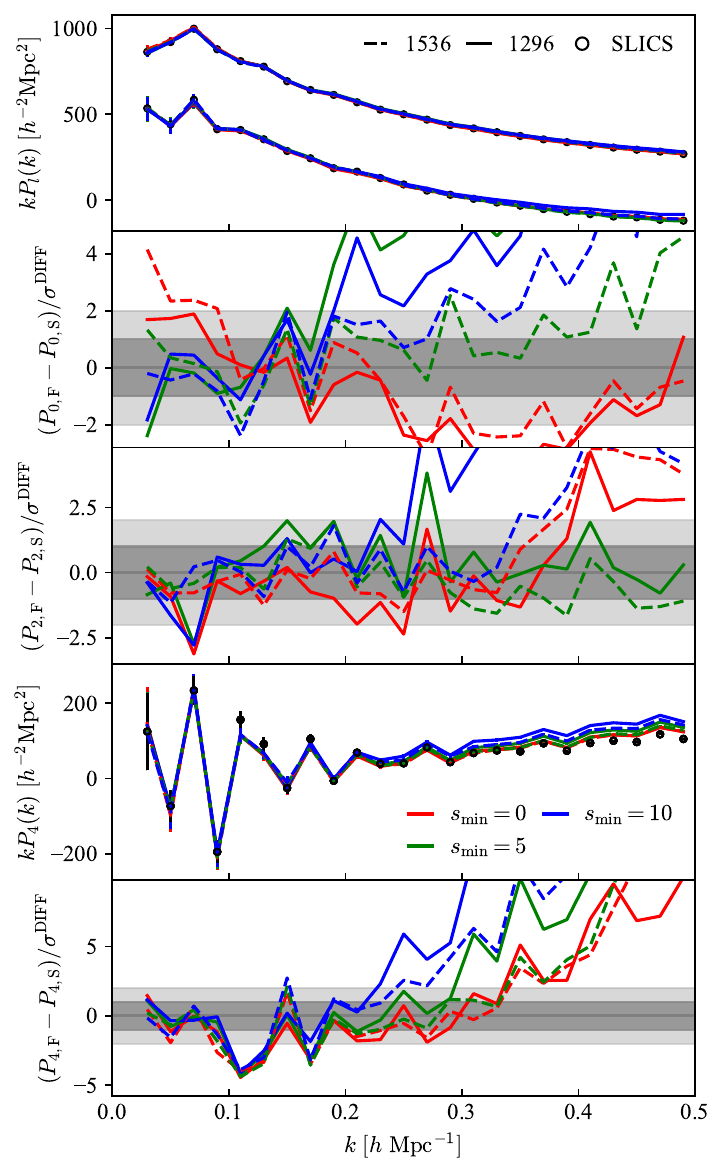}
    \caption{Same as Figure~\ref{fig:1296_1536_DIFF_PK_kmax_05_04_03}, but the fitting is done on the monopole and quadrupole of the 2PCF.}
    \label{fig:1296_1536_DIFF_CF_smin_0_5_10}
\end{figure*}

Figure~\ref{fig:1296_1536_DIFF_CF_smin_0_5_10} presents the monopole, quadrupole and hexadecapole of the 2PCF computed for $s \in [0, 200]\,\su$ as well as the tensions between the \textsc{FastPM} and \textsc{SLICS}. The \textsc{FastPM} clustering typically falls within $2 \sigma$ of the reference for scales larger than $50\,\su$ and is largely unaffected by the fitting scenario.
However, the HR monopoles are consistently closer to the reference than LR monopoles by approximately $0.5\sigma$ at scales larger than $\approx150\,\su$.

Including the smallest scales (Large interval) in the HOD fitting, we observe a 1 to $2\sigma$ agreement with the reference for $s < 10\,\su$ in both the monopole and quadrupole. However, at intermediate scales $s \in [10, 50]\,\su$, the monopole is significantly biased, exhibiting a deviation of $3 \sigma$. In contrast, for the Medium and Small scenarios, we notice that the tensions for the monopole and quadrupole at intermediate scales drop to $1\sigma$, while the smallest scales can get biased by more than $3 \sigma$. Nevertheless, they match better the reference than the power spectrum HOD fitting case.
Lastly, the hexadecapole does not depend on the resolution nor the fitting range and is strongly biased for $s<60\,\su$, showing no improvement compared to the power spectrum fitting.

As in the previous subsection, we test the clustering statistics of the best-fitting \textsc{FastPM} boxes that were not included in the HOD fitting, i.e. the power spectrum in the Figure~\ref{fig:1296_1536_DIFF_CF_smin_0_5_10}. The first observation is that these \textsc{FastPM} power spectra do not fit as well the reference as the ones from Figure~\ref{fig:1296_1536_DIFF_PK_kmax_05_04_03}. On one hand, for the HR case and Medium and Small fitting intervals a $\pm1\sigma$ matching is possible up to $k=0.4\,\ku$ and $k=0.3\,\ku$, respectively. On the other hand, the LR \textsc{FastPM} allows a good matching up to $k\approx0.2\,\ku$ for the same fitting intervals.
While the $s_\mathrm{min}=0$ case has a good matching quadrupole up to $k\approx0.4\,\ku$, its monopole follows similar trend to the 2PCF monopole, i.e. the intermediate scales $k \in [0.25, 0.4]\,\ku$ are biased and the rest are mostly within $2\sigma$ deviation.
Lastly, the hexadecapole is within $\pm2\sigma$ up to $k\approx0.3\,\ku$ for the Large fitting interval and up to $k\approx0.2\,\ku$  for the other cases.

A quantitative evidence that directly fitting the 2PCF yields superior matching of the 2PCF compared to fitting the power spectrum is displayed in Figure~\ref{fig:reduced_chi2}. The majority of the $\chi^2_\nu$ values in the lower-right panel are lower compared to those in the upper-right panel. Furthermore, fitting on the Small interval ($s_\mathrm{min} = 10$), the $\chi^2_\nu\approx1$, indicating that the 2PCF is in good agreement with the \textsc{SLICS} reference above a certain minimum separation. 
The almost constant $\chi^2_\nu$ for the Large fitting interval in the lower-right panel of Figure~\ref{fig:reduced_chi2} is explained by the discrepancy at the intermediate scales of the monopole for the Large fitting interval in Figure~\ref{fig:1296_1536_DIFF_CF_smin_0_5_10}.
Lastly, as in the previous fitting scenario, the HR \textsc{FastPM} generally provides a lower $\chi^2_\nu$ than the LR. In contrast, only the HR simulations can provide a $\chi^2_\nu<2$ to both the 2PCF and the power spectra, and only when fitting with the Medium and Small intervals to 2PCF. Although not shown in the aforementioned figure, it is important to note that $\chi^2_\nu[P(0.02, 0.2)] = 2.4$ for the 2PCF Small interval LR case.

\subsection{Bi-spectrum comparison}

The two-point clustering covariance matrix is directly related to the tri-spectrum, i.e. the four-point clustering, however this is difficult to tune. Nonetheless, \citet{2018MNRAS.480.2535B} have shown that for similar two-point clustering, the corresponding covariance matrices -- at scales of $s < 40\,\su$ -- are sensitive to changes in the bi-spectrum. Consequently, even though we do not include the bi-spectrum into the HOD fitting, we aim to understand its behaviour when incorporating various scales of the two-point clustering in the HOD fitting.

\begin{figure*}
	\includegraphics[width=\columnwidth]{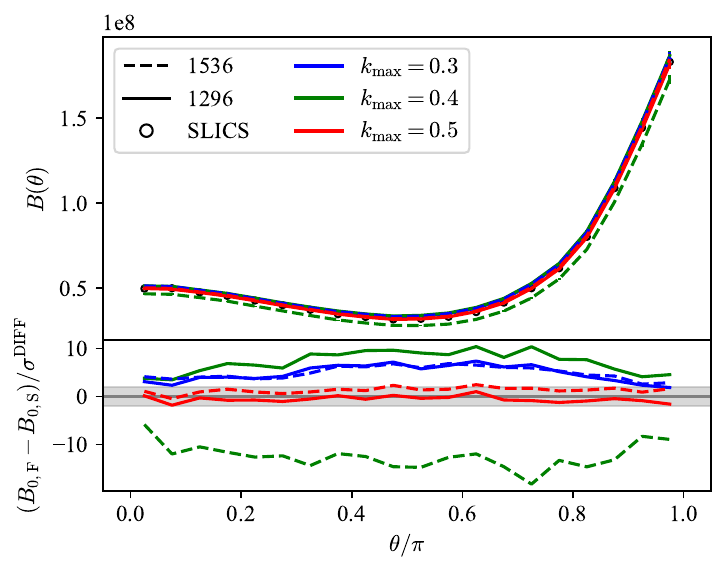}
    \includegraphics[width=\columnwidth]{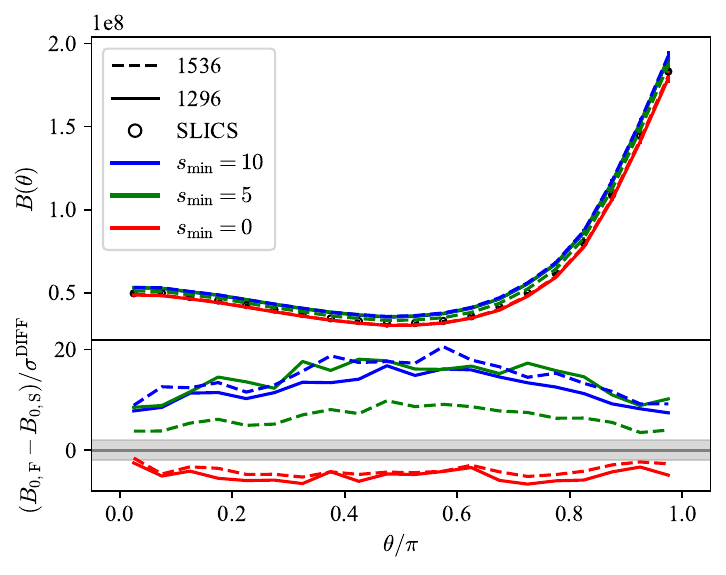}
    \caption{A comparison between the average \textsc{SLICS} bi-spectrum and the average \textsc{FastPM} bi-spectrum. The averages are computed from the 20 realisations used during the HOD fitting. The left panel shows the results from fitting the power spectrum, as in Figure~\ref{fig:1296_1536_DIFF_PK_kmax_05_04_03}. The right panel displays the results from fitting the 2PCF, as in Figure~\ref{fig:1296_1536_DIFF_CF_smin_0_5_10}. The shaded area denotes $\pm2\sigma$ deviation. The bi-spectra are computed for $k_1 = 0.1 \pm 0.05 \,\ku$ and $k_2 = 0.2 \pm 0.05\,\ku$, with the $\theta$ angle between $\mathbf{k_1}$ and $\mathbf{k_2}$ varying from 0 to $\pi$.}
    \label{fig:1296_1536_bispec_20}
\end{figure*}

Figure~\ref{fig:1296_1536_bispec_20} compares the average bi-spectrum of the 20 best-fitting \textsc{FastPM} boxes with the one computed on the corresponding \textsc{SLICS} boxes, for $2\,k_1 = k_2 = 0.2\,\ku$ configuration as done by \citet{2018MNRAS.480.2535B}. These scales are chosen because they are sensitive to BAO and RSD, the primary scientific case of DESI. It is evident that by increasing the fitting range to include smaller scales, the \textsc{FastPM} bi-spectrum changes to the extent that for $k_\mathrm{max}=0.5$, the tension ranges from 1 to $2\sigma$. In contrast, when fitting the 2PCF the resulting bi-spectrum is more biased, i.e. the lowest deviation is $\approx5\sigma$, for $s_\mathrm{min}=0$ case. 
In addition, we have checked multiple configurations of the bi-spectrum with $k_i\in[0.01, 0.21]\,\ku$ and $i=1$ or $2$ and observed that for $k_\mathrm{max}=0.5$ and $s_\mathrm{min}=0$, the overall agreement between the \textsc{FastPM} and \textsc{SLICS} is around $1$ and $5$ per cent, respectively. Lastly, there is no significant improvement in terms of the goodness-of-fit between the HR and LR.

In the previous sections, we compare the HR and LR \textsc{FastPM} with \textsc{SLICS} using the two-point clustering of the 20 cubic mocks included in the HOD fitting. The HR simulations perform better than LR to model the extremely non-linear scales, such as $k\approx0.5\,\ku$, $s\approx0\,\su$. In contrast, at mildly non-linear scales ($k\approx0.3\,\ku$, $s\approx10\,\su$) that are more relevant to BAO and RSD analyses \citep[e.g. ][]{2020MNRAS.499.5527T, 2021MNRAS.501.5616D}, LR and HR show similar performance. Moreover, Figure~\ref{fig:1296_1536_bispec_20} suggests that the bi-spectrum does not depend strongly on the resolution. Nevertheless, the computing cost of HR is significantly higher than for LR and given the small difference in precision, we argue it is optimal to use LR \textsc{FastPM} for further analyses.

\begin{figure}
	\includegraphics[width=\columnwidth]{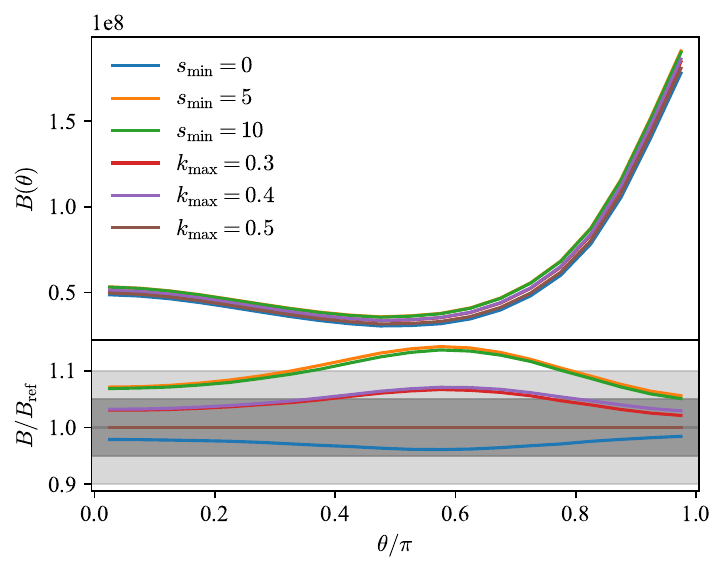}
    \caption{Average bi-spectra computed using 778 LR \textsc{FastPM} realisations for different HOD fitting cases, see Table~\ref{tab:fitting_intervals_n_bins}. The reference for the bi-spectra ratios is the average bi-spectrum computed for the $k_\mathrm{max}=0.5$ case. The bi-spectra are computed for $k_1 = 0.1 \pm 0.05\,\ku$ and $k_2 = 0.2 \pm 0.05\,\ku$, with the $\theta$ angle between $\mathbf{k_1}$ and $\mathbf{k_2}$ varying from 0 to $\pi$.}
    \label{fig:bispec_1296_780}
\end{figure}

Furthermore, in Figure~\ref{fig:bispec_1296_780}, we compare the average bi-spectra -- computed from 778 LR \textsc{FastPM} realisations -- corresponding to the six HOD fitting cases, see Table~\ref{tab:fitting_intervals_n_bins}. In this and the next figures, we choose the $k_\mathrm{max}=0.5$ case as reference because:
\begin{enumerate}
    \item Figure~\ref{fig:reduced_chi2} shows that the best-fitting power spectrum provides $\chi^2_\nu\approx1$;
    \item Figure~\ref{fig:1296_1536_bispec_20} implies that the corresponding bi-spectrum is the closest to the \textsc{SLICS} reference. 
\end{enumerate}

One can notice that $s_\mathrm{min}=0$ bi-spectrum is at most 5 per cent different than the reference, while the rest can reach 15 per cent discrepancies. The $k_\mathrm{max}=0.3$ and $k_\mathrm{max}=0.4$ cases are 1 to 2 per cent different from each other and similarly for $s_\mathrm{min}=5$ and $s_\mathrm{min}=10$.

\subsection{Covariance matrix comparison}

\begin{figure}
	\includegraphics[width=\columnwidth]{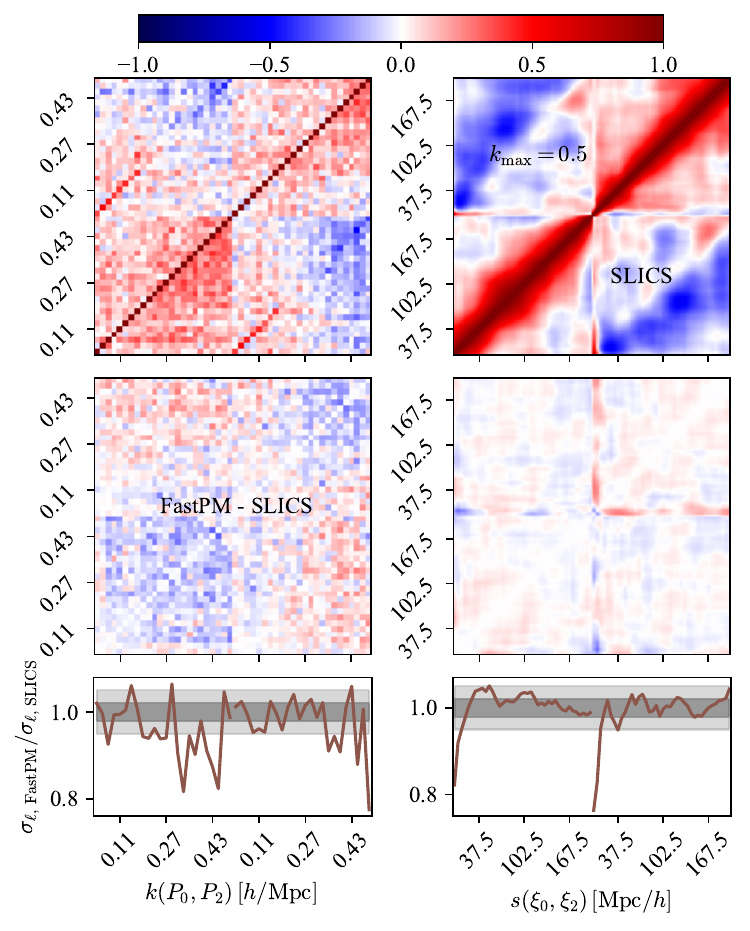}
    \caption{The correlation matrices and the standard deviations computed using 123 realisations of the monopole and quadrupole of the power spectrum (left) and 2PCF (right). In the first row, the upper triangular matrices display the correlation matrices of the \textsc{FastPM} $k_\mathrm{max}=0.5$ case, while the lower triangular ones show the \textsc{SLICS} correlation matrices. The second row of panels contains the differences between the \textsc{SLICS} and \textsc{FastPM} correlation matrices. The third row of panels illustrate the ratios between the standard deviations. The shaded regions denote two and five per cent limits.}
    \label{fig:covariance_fastpm_slics_123}
\end{figure}

Having studied the behaviour of the bi-spectra with respect to the scales introduced into the HOD fitting, we now want to understand the resulting impact on the two-point clustering covariance matrices.
Firstly, Figure~\ref{fig:covariance_fastpm_slics_123} compares the $k_\mathrm{max}=0.5$ and \textsc{SLICS} correlation matrices and standard deviations. Given the limited number of realisations used to estimate the correlation matrices, it is difficult to assess the importance of the differences. Nevertheless, there seems to be more significant differences at very small scales for the 2PCF and important differences at almost all scales for the power spectrum. Numerically, the standard deviations deviate by more than 7 to 10 per cent from the \textsc{SLICS} case for $s<10\,\su$ and $k>0.35\,\ku$. A more detailed comparison is performed by \citet{DESI_MOCK_CHALLENGE_I}.

Secondly, we compare the correlation matrices and standard deviations obtained from 778 \textsc{FastPM} realisations, of the six HOD fitting cases, where $k_\mathrm{max}=0.5$ is the reference.

\subsubsection{Power spectrum covariance matrix}

\begin{figure}
	\includegraphics[width=\columnwidth]{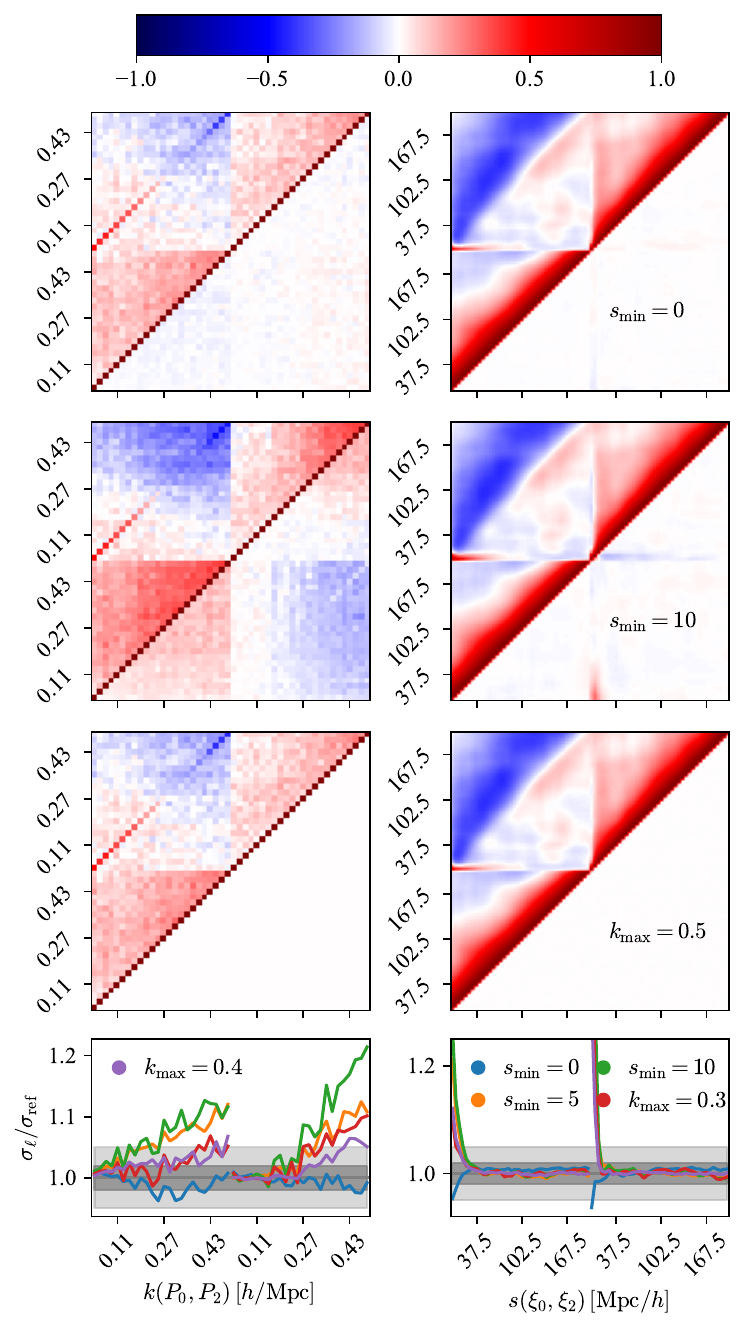}
    \caption{The correlation matrices and the standard deviations computed using 778 realisations of the monopole and quadrupole of the power spectrum (left) and 2PCF (right). Given the similarity between some correlation matrices, only three out of six different HOD fitting cases -- see Table~\ref{tab:fitting_intervals_n_bins} -- are presented here. However, all cases can be found in Appendix~\ref{sec:appendix_cov_mat_comp}. The reference case corresponds to $k_\mathrm{max}=0.5$. The upper triangular matrices display the correlation matrices, while the lower triangular ones show the differences between the correlation matrices and the reference one. The bottom panels illustrate the ratios between the standard deviations. The shaded regions denote two and five per cent limits.}
    \label{fig:correlation_together_2pcf_pspec_1296_780}
\end{figure}

Figure~\ref{fig:correlation_together_2pcf_pspec_1296_780} presents the correlation matrices and the corresponding standard deviations $\sigma_\ell$ for the monopole and quadrupole of the power spectrum. The following pairs ($k_\mathrm{max}=0.4$, $k_\mathrm{max}=0.3$), ($s_\mathrm{min}=5$, $s_\mathrm{min}=10$) and ($s_\mathrm{min}=0$, $k_\mathrm{max}=0.5$) have very similar correlation matrices, thus we only show three cases. However, we introduce all of them in Appendix~\ref{sec:appendix_cov_mat_comp}.

The similarity to the reference correlation matrix diminishes in the following order:  $s_\mathrm{min}=0$, $k_\mathrm{max}=0.4$, $k_\mathrm{max}=0.3$, $s_\mathrm{min}=5$ and $s_\mathrm{min}=10$. However, for the largest scales of the quadrupole ($k<0.15\,\ku$), the correlation coefficients are practically the same for all cases. 

The standard deviations in the lowest panels show similar trends.  The $s_\mathrm{min}=0$ case is within two per cent of the reference case. The $k_\mathrm{max}=0.4$, $k_\mathrm{max}=0.3$ cases overestimate the $\sigma_\ell(k)$ by approximately two per cent for $k<0.27\,\ku$ and by $\approx 5$ per cent for smaller scales. Nevertheless, these two cases are consistent with each other within one to two per cent. In contrast, $s_\mathrm{min}=5$ and $s_\mathrm{min}=10$ can overestimate the $\sigma_\ell(k)$ by $\approx 2$ to 5 per cent for $k<0.27\,\ku$ and by 10 to 20 per cent for smaller scales. These two cases are also consistent with each other for most scales, except for the quadrupole $k>0.3\,\ku$.
These findings are in agreement with the trends observed in the bi-spectrum comparison in Figure~\ref{fig:bispec_1296_780}.

\begin{figure}
	\includegraphics[width=\columnwidth]{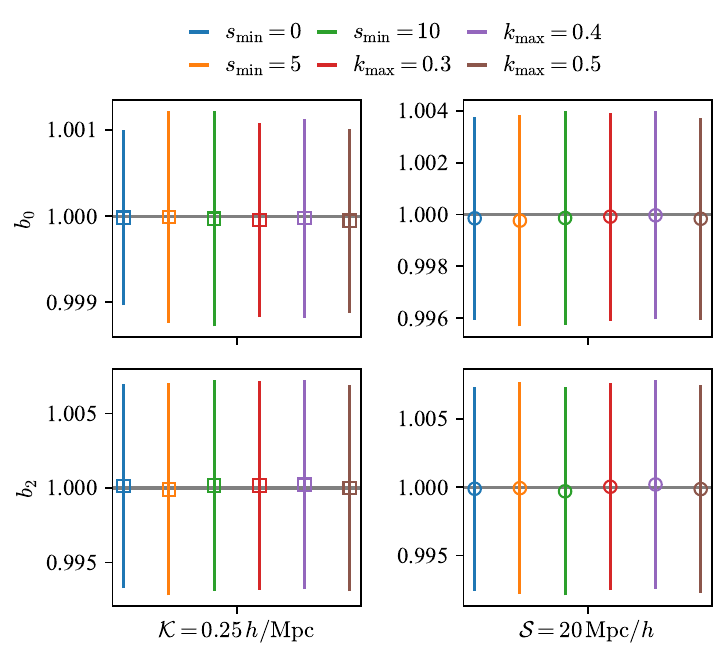}
    \caption{The average of the 123 fitting parameters ($b_\ell$) obtained from 123 \textsc{SLICS} clustering realisations, as described in Section~\ref{sec:cov_mat_constrain_power}. The measurements performed on the power spectra with $k\in[0.02, \mathcal{K}]\,\ku$ are show on the left, while those based on the 2PCF with $s\in[\mathcal{S}, 200]\,\su$ are depicted on the right. The error bars are computed as the average of 123 $\sigma_{b_\ell}$, divided by $\sqrt{123}$. Here, $\sigma_{b_\ell}$ represents the standard deviation of the $b_\ell$ posterior distribution. The colours correspond to the different \textsc{FastPM} covariance matrices illustrated in Figure~\ref{fig:correlation_together_2pcf_pspec_1296_780}. Due to the similar results, only one value for $\mathcal{K}$ and $\mathcal{S}$ are shown here. However, all tested values are presented in Appendix~\ref{sec:appendix_cov_mat_comp}} 
    \label{fig:together_pspec_2pcf_fit_avg_b0_b2}
\end{figure}

\begin{figure*}
	\includegraphics[width=\textwidth]{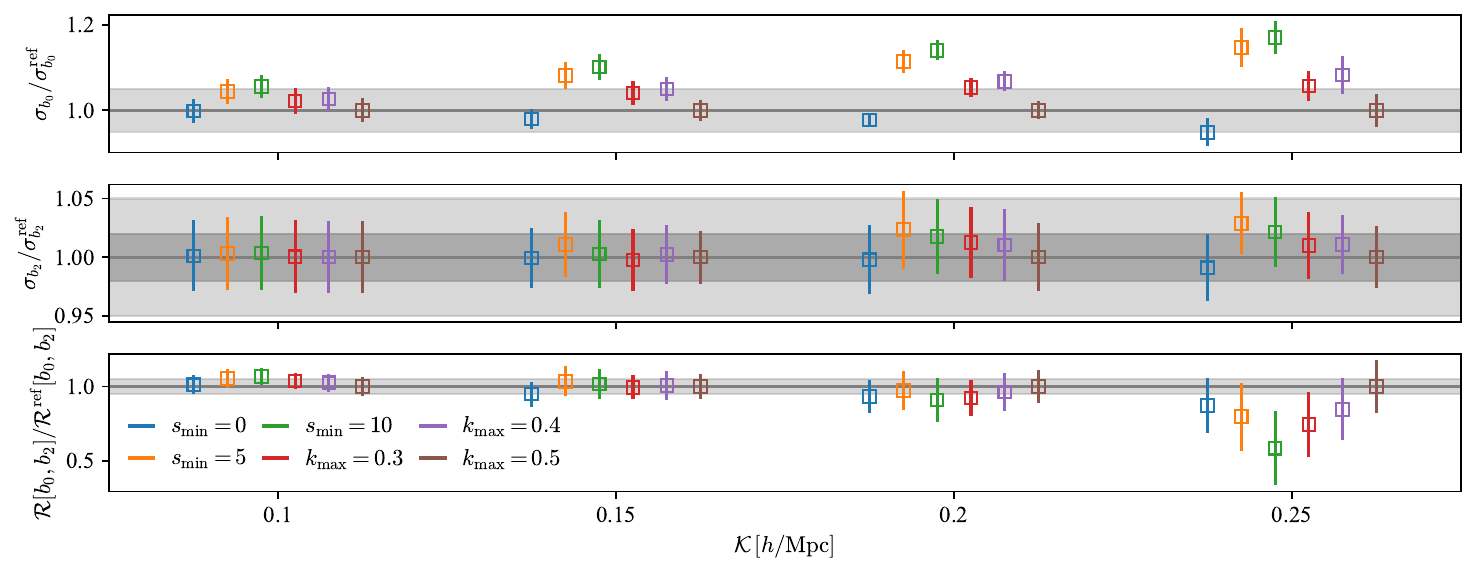}
    \caption{The averages of 123 $\sigma_{b_\ell}$ and 123 $\mathcal{R}[b_0, b_2]$ -- the standard deviation of the $b_\ell$ posterior distribution and the covariance between $b_0$ and $b_2$, respectively, detailed in Section~\ref{sec:cov_mat_constrain_power} -- obtained from 123 \textsc{SLICS} power spectra fitted on $k\in[0.02, \mathcal{K}]\,\ku$. In order to estimate the error bars, we split the 778 \textsc{FastPM} realisations in six distinct sets of 123 realisations and compute for each set $u$ a covariance matrix $\Sigma^u_\mathrm{123, \textsc{FastPM}}$ with which we fit the 123 \textsc{SLICS} clustering realisations. Having obtained 123 values of $\sigma_{b_\ell}^u$ per set, we compute their average $\Bar{\sigma}^u$. Finally, the error bars are the standard deviation of the six $\Bar{\sigma}^u$ divided by $\sqrt{6}$. The different colours stand for the different \textsc{FastPM} covariance matrices exhibited in Figure~\ref{fig:correlation_together_2pcf_pspec_1296_780}. The $k_\mathrm{max}=0.5$ represents the reference, i.e. all values ($\sigma_{b_\ell}$ and its error bars) are scaled by the $\sigma_{b_\ell}$ corresponding to $k_\mathrm{max}=0.5$ case. This is why all brown squares have the value of one. The shaded areas delineate the 2 and 5 per cent regions with respect to the reference.}
    \label{fig:pspec_fit_std_b0_b2}
\end{figure*}

In order to quantify the differences between the covariance matrices we adopt the method described in Section~\ref{sec:cov_mat_constrain_power} and thus obtain the results displayed in Figures~\ref{fig:together_pspec_2pcf_fit_avg_b0_b2} and \ref{fig:pspec_fit_std_b0_b2}. The first one reveals that none of the six covariance matrices bias the two fitting parameters, regardless of the fitting range. We only present here the results of one fitting range, however all cases can be found in Appendix~\ref{sec:appendix_cov_mat_comp}.

Examining the uncertainty on $b_0$ in Figure~\ref{fig:pspec_fit_std_b0_b2}, we observe that including the smaller scales the discrepancy between the error estimates of the six covariance matrices increases, as we expect from Figure~\ref{fig:correlation_together_2pcf_pspec_1296_780}, reaching a maximum of $\approx 20$ per cent larger error estimation for the $s_\mathrm{min}=10$ covariance at $\mathcal{K}=0.25\,\ku$. Moreover, each of the following pairs ($k_\mathrm{max}=0.4$, $k_\mathrm{max}=0.3$), ($s_\mathrm{min}=5$, $s_\mathrm{min}=10$) and ($s_\mathrm{min}=0$, $k_\mathrm{max}=0.5$) provide coherent estimations of the uncertainty, which is consistent with the observations on the correlation matrices and standard deviations. Lastly, a five per cent consensus between all six covariance matrices is achieved when we fit the power spectra on the $k\in[0.02, 0.1]\,\ku$.

The agreement between covariance matrices on $\sigma_{b_2}$ is much better than $\sigma_{b_0}$. Given the error bars, the six methods estimate the uncertainty with a two per cent tolerance with each other for all $\mathcal{K}$ values.

Finally, all six covariance matrices provide values of $\mathcal{R}[b_0,b_2]$  that are consistent at the level of 5 per cent, given the error bars and up to $\mathcal{K}=0.2\,\ku$. For $\mathcal{K}=0.25\,\ku$, the largest discrepancy is shown by $s_\mathrm{min}=10$ case which underestimates the value of $\mathcal{R}[b_0,b_2]$ by almost 50 per cent. The other cases underestimate $\mathcal{R}[b_0,b_2]$ by 10 to 20 per cent.

\subsubsection{2PCF covariance matrix}

Comparing the correlation matrices obtained from 778 2PCF in Figure~\ref{fig:correlation_together_2pcf_pspec_1296_780}, one can observe that the largest differences occur at the smallest scales $s<30\,\su$. Similarly to the power spectrum correlation matrices, the same pairs of cases show resembling behaviours at all scales. Equivalent qualitative comments can be made about the ratios of the standard deviations. Nonetheless, all cases are within $\approx2$ per cent from each other for $s>30\,\su$, while at smaller scales, the differences can get larger than $\approx20$ per cent.

\begin{figure*}
	\includegraphics[width=\textwidth]{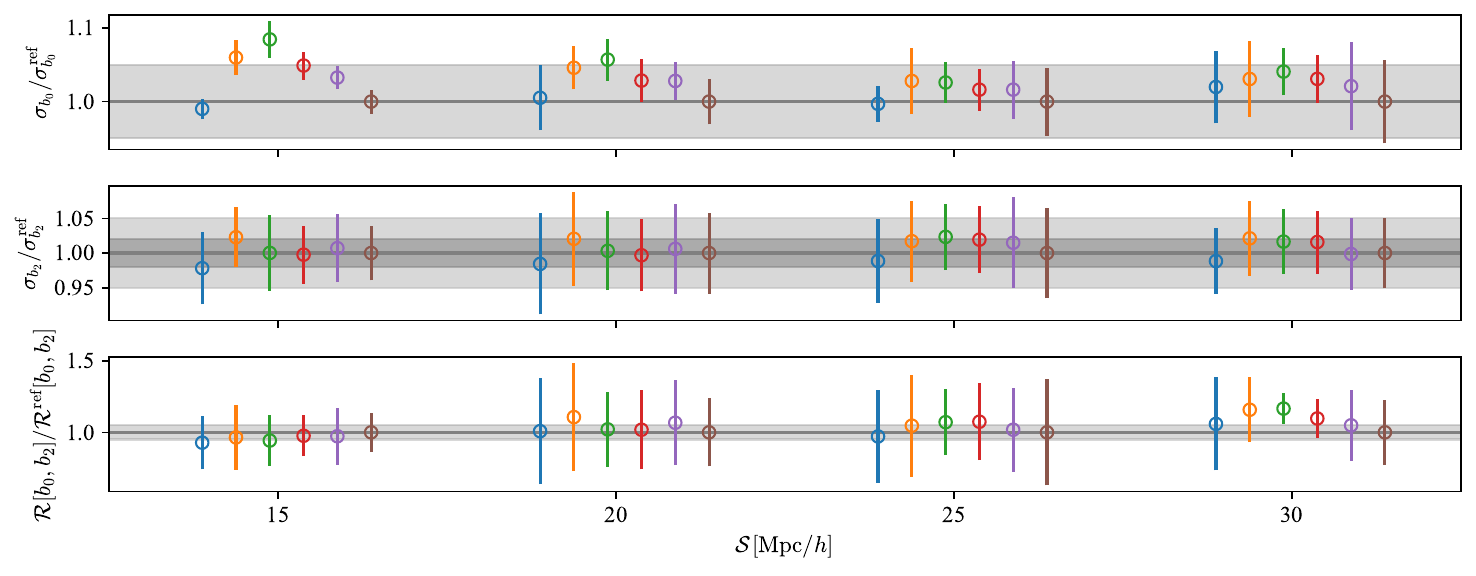}
    \caption{Same as Figure~\ref{fig:pspec_fit_std_b0_b2}, but the fitting is performed on 123 \textsc{SLICS} 2PCF and $s\in[\mathcal{S}, 200]\,\su$.}
    \label{fig:2pcf_fit_std_b0_b2}
\end{figure*}

Following the method described in Section~\ref{sec:cov_mat_constrain_power}, we obtain the results shown in Figures~\ref{fig:together_pspec_2pcf_fit_avg_b0_b2} and \ref{fig:2pcf_fit_std_b0_b2}. The first figure proves that all six covariance matrices provide unbiased measurements of $b_\ell$ parameters.

Resembling the power spectrum fitting case, the six estimations of $\sigma_{b_0}$ in Figure~\ref{fig:2pcf_fit_std_b0_b2} are in better agreement when the smallest scales are not included in the 2PCF fitting, however for $\mathcal{S}\geq20\,\su$ they are all within $\approx5$ per cent from each other. The largest discrepancy is around 10 per cent and occurs between $s_\mathrm{min}=0$ and $s_\mathrm{min}=10$ for $\mathcal{S}=15\,\su$. The values of $\sigma_{b_2}$ are all consistent within $\approx2$ per cent, given the error bars. Interestingly, including the smaller scales, the $\mathcal{R}[b_0, b_2]$ values are more coherent, such that all discrepancies are within five per cent, given the error bars and for $\mathcal{S}<30\,\su$. In contrast, when $\mathcal{S}=30\,\su$ the $s_\mathrm{min}=10$ and $s_\mathrm{min}=5$ provide values $\mathcal{R}[b_0, b_2]$ that are approximately ten per cent larger than the reference, but nevertheless consistent within the error bars.

The covariance matrix obtained through fitting the power spectrum or the 2PCF can be different on highly non-linear scales, therefore an analysis using such scales needs to be more careful with the covariance matrix estimation. In contrast, in the quasi-linear regime ($s \gtrsim 20\,\su$; $k \lesssim 0.2\,\ku$) the differences are not statistically significant (see e.g. $s_\mathrm{min}=0$ and $k_\mathrm{max}=0.5$ cases) , thus one can use either space to obtain the covariance matrices and apply them to both analyses.

\section{Conclusions}
\label{sec:conclusion}

We have implemented an HOD model to assign galaxies on the \textsc{FastPM} halo cubic mocks, such that the resulting clustering -- monopole and quadrupole -- matches the \textsc{SLICS} reference one. In order to remove the cosmic variance, we have used 20 \textsc{SLICS} galaxy catalogues  and 20 halo \textsc{FastPM} mocks (low resolution or high resolution) that share the initial conditions with the \textsc{SLICS} simulations. Given the shared white noise, the standard covariance matrix is obsolete, thus we have performed a two-steps HOD fitting:
\begin{enumerate}
    \item use a simple diagonal covariance matrix to get Initial-Guess best-fitting \textsc{FastPM} galaxy mocks;
    \item compute the covariance matrix of the 123 realisations of the difference between the IG-\textsc{FastPM} and the \textsc{SLICS} clustering, and use it to perform the final HOD fitting. 
\end{enumerate}
The final HOD fitting has been performed on three different fitting ranges for both power spectrum $k\in[0.02, k_\mathrm{max}]\,\ku$ and 2PCF $s\in[s_\mathrm{min}, 50]\,\su$: ($k_\mathrm{max}=0.5$, $k_\mathrm{max}=0.4$, $k_\mathrm{max}=0.3$) and ($s_\mathrm{min}=0$, $s_\mathrm{min}=5$, $s_\mathrm{min}=10$), respectively.

On one hand, the HR \textsc{FastPM} generally performs better than the LR at modelling the \textsc{SLICS} clustering. On the other hand, LR is also able to provide a $\chi^2_\nu\approx1$ for $k_\mathrm{max}=0.5$, $k_\mathrm{max}=0.4$, $k_\mathrm{max}=0.3$ and $s_\mathrm{min}=10$. The $k_\mathrm{max}=0.5$ case is one of the most valuable as it additionally offers $2\sigma$ matching:
\begin{enumerate}
    \item power spectrum hexadecapole for $k<0.4\,\ku$;
    \item 2PCF monopole and quadrupole for $s>10\,\su$;
    \item bi-spectrum. 
\end{enumerate}
Nevertheless, fitting the 2PCF with $s_\mathrm{min}=10$, produce a $1\sigma$ matching power spectrum monopole and quadrupole for $k\lesssim 0.2$, but a strongly biased bi-spectrum. In a similar way as the power spectrum, one must include the smallest scales to better reproduce the \textsc{SLICS} bi-spectrum, i.e. for $s_\mathrm{min}=0$ the bi-spectrum tension drops from $20\sigma$ to $5\sigma$. 
As a general remark, the power spectrum hexadecapoles can be slightly tuned by changing the values of $k_\mathrm{max}$ or $s_\mathrm{min}$, but the 2PCF hexadecapole is practically independent on the fitting range.

The fact that there is no HOD fitting case that provides both power spectrum and 2PCF $1\sigma$ matching with the reference might raise the question whether one needs two sets of catalogues for the DESI analysis. Therefore, it could be interesting for future studies to perform a joint fitting of both Fourier and Configuration clustering statistics to test for possible improvements in modelling non-linear scales. Future studies could test also the necessary level of agreement between the \textsc{FastPM} and the full $N$-body reference in order to have a similar behaviour when observational systematic effects are included. Nonetheless, for the purpose of this article, in the second part of the study, we have exposed and compared the resulting covariance matrices. 

Firstly, we have briefly shown with the 123 \textsc{FastPM} realisations -- that share the initial conditions with \textsc{SLICS} -- that the standard deviations of the $k_\mathrm{max}=0.5$ case agree within five per cent with the \textsc{SLICS} reference case for $s>10\,\su$ and $k<0.35\,\ku$. Secondly, we have focused on the 778 LR \textsc{FastPM} realisations corresponding to the six best-fitting cases, where $k_\mathrm{max}=0.5$ is considered the reference. 

Since \citet{2018MNRAS.480.2535B} have shown that the bi-spectrum computed with a configuration that includes the BAO and RSD effects -- i.e. $2 \times k_1 = k_2 = 0.2 \,\ku$ -- affects the covariance matrix, we have compared the six bi-spectra and checked how this comparison reflects into the two-point clustering covariance matrices. Lastly, we have examined the constraining power of these covariance matrices using a simplified clustering model with two scaling parameters, i.e. $b_0$ and $b_2$ for the monopole and quadrupole. We have focused on fitting intervals similar to the ones used in standard BAO and RSD analyses i.e. $\mathcal{K} \lesssim 0.20\,\ku$ and $\mathcal{S} \gtrsim20\,\su$ \citep[e.g. ][]{2020MNRAS.499.5527T, 2021MNRAS.501.5616D}.

The $s_\mathrm{min}=0$ bi-spectrum is at most five per cent different than the $k_\mathrm{max}=0.5$, while the other cases can reach a discrepancy of 15 per cent. However, each of these pairs ($k_\mathrm{max}=0.4$, $k_\mathrm{max}=0.3$), ($s_\mathrm{min}=5$, $s_\mathrm{min}=10$) yield similar bi-spectra. 
These observations are in a good agreement with a qualitative description of the shown correlation matrices and the standard deviations.

Quantitatively, the power spectrum standard deviations of $s_\mathrm{min}=0$ and ($k_\mathrm{max}=0.4$, $k_\mathrm{max}=0.3$) are within two percent from the reference for $k<0.5\,\ku$ and $k<0.27\,\ku$, respectively. Furthermore, the 2PCF standard deviations of all cases are within two percent from the reference for $s>30\,\su$.

Using the simplified clustering model, $b_0$ and $b_2$ are measured accurately for both power spectrum and 2PCF using all six covariance matrices. The six estimations of $\sigma_{b_0}$ from the power spectrum fitting up to $\mathcal{K}=0.20\,\ku$ are scattered within at most 20 per cent from the reference, whereas the values of $\sigma_{b_2}$ are within two per cent agreement, given the error bars. Lastly, the covariances between $b_0$ and $b_2$ are scattered within 5 per cent from the reference.

In contrast, the estimations of $\sigma_{b_0}$ from the 2PCF fitting down to $\mathcal{S} = 20\,\su$ are found within five per cent from each other. Similarly to the power spectrum case, the $\sigma_{b_2}$ values agree at the level of two per cent. Given the error bars, the covariances between $b_0$ and $b_2$ are consistent at the level of five per cent.

Lastly, analysing individual cases in the following pairs of HOD fitting cases: ($s_\mathrm{min}=0$, $k_\mathrm{max}=0.5$), ($k_\mathrm{max}=0.4$, $k_\mathrm{max}=0.3$), ($s_\mathrm{min}=5$, $s_\mathrm{min}=10$), one can observe that each pair provides similar uncertainty estimation $\sigma_{b_\ell}$, irrespective of the level of matching between the \textsc{FastPM} and \textsc{SLICS} two-point clustering. In order words, it seems that the estimations of the uncertainties are closer when the bi-spectra -- computed for $2 \times k_1 = k_2 = 0.2 \,\ku$ -- are more similar. The most striking case is the 2PCF for $s_\mathrm{min}=0$ which is more than $2\sigma$ different than the $k_\mathrm{max}=0.5$ 2PCF, for $s<40\,\su$, but the 2PCF standard deviations and the $\sigma_{b_\ell}$ estimations are practically identical.

In conclusion, one can use an HOD model on the low resolution \textsc{FastPM} halo catalogues to tune the galaxy clustering such that it matches the \textsc{SLICS} reference down to certain minimum scales. Additionally, the HOD fitting intervals can have an impact on the final \textsc{FastPM} based covariances. This influence is observed as a scatter in the uncertainty estimation of up to 20 per cent for power spectrum and five per cent for 2PCF at the scales interesting for BAO and RSD analyses. Nevertheless, more accurate analyses could be performed in the future using actual BAO and RSD models and a larger sample of mocks, such as \textsc{AbacusSummit}.


\section*{Acknowledgements}
AV, CZ, DFS acknowledge support from the Swiss National Science Foundation (SNF) "Cosmology with 3D Maps of the Universe" research grant, 200020\_175751 and 200020\_207379.
SA acknowledge support of the Department of Atomic Energy, Government of india, under project no. 12-R\&D-TFR-5.02-0200. SA is partially supported by the European Research Council through the COSFORM Research Grant (\#670193) and STFC consolidated grant no. RA5496. The authors are thankful for the constructive comments of Joseph DeRose that have helped improved the clarity and quality of our article.

This material is based upon work supported by the U.S. Department of Energy (DOE), Office of Science, Office of High-Energy Physics, under Contract No. DE–AC02–05CH11231, and by the National Energy Research Scientific Computing Center, a DOE Office of Science User Facility under the same contract. Additional support for DESI was provided by the U.S. National Science Foundation (NSF), Division of Astronomical Sciences under Contract No. AST-0950945 to the NSF’s National Optical-Infrared Astronomy Research Laboratory; the Science and Technology Facilities Council of the United Kingdom; the Gordon and Betty Moore Foundation; the Heising-Simons Foundation; the French Alternative Energies and Atomic Energy Commission (CEA); the National Council of Science and Technology of Mexico (CONACYT); the Ministry of Science and Innovation of Spain (MICINN), and by the DESI Member Institutions: \url{https://www.desi.lbl.gov/collaborating-institutions}. Any opinions, findings, and conclusions or recommendations expressed in this material are those of the author(s) and do not necessarily reflect the views of the U. S. National Science Foundation, the U. S. Department of Energy, or any of the listed funding agencies.

The authors are honored to be permitted to conduct scientific research on Iolkam Du’ag (Kitt Peak), a mountain with particular significance to the Tohono O’odham Nation.

\section*{Data Availability}

The \textsc{SLICS} and \textsc{FastPM} boxes used in this study can be provided upon reasonable request to authors. The clustering measurements can be found on Zenodo \url{https://doi.org/10.5281/zenodo.8185822}.



\bibliographystyle{mnras}
\bibliography{refs} 

\begin{thebibliography}{}
\makeatletter
\relax
\def\mn@urlcharsother{\let\do\@makeother \do\$\do\&\do\#\do\^\do\_\do\%\do\~}
\def\mn@doi{\begingroup\mn@urlcharsother \@ifnextchar [ {\mn@doi@} {\mn@doi@[]}}
\def\mn@doi@[#1]#2{\def\@tempa{#1}\ifx\@tempa\@empty \href {http://dx.doi.org/#2} {doi:#2}\else \href {http://dx.doi.org/#2} {#1}\fi \endgroup}
\def\mn@eprint#1#2{\mn@eprint@#1:#2::\@nil}
\def\mn@eprint@arXiv#1{\href {http://arxiv.org/abs/#1} {{\tt arXiv:#1}}}
\def\mn@eprint@dblp#1{\href {http://dblp.uni-trier.de/rec/bibtex/#1.xml} {dblp:#1}}
\def\mn@eprint@#1:#2:#3:#4\@nil{\def\@tempa {#1}\def\@tempb {#2}\def\@tempc {#3}\ifx \@tempc \@empty \let \@tempc \@tempb \let \@tempb \@tempa \fi \ifx \@tempb \@empty \def\@tempb {arXiv}\fi \@ifundefined {mn@eprint@\@tempb}{\@tempb:\@tempc}{\expandafter \expandafter \csname mn@eprint@\@tempb\endcsname \expandafter{\@tempc}}}

\bibitem[\protect\citeauthoryear{{Alam} et~al.,}{{Alam} et~al.}{2017}]{2017MNRAS.470.2617A}
{Alam} S.,  et~al., 2017, \mn@doi [\mnras] {10.1093/mnras/stx721}, \href {https://ui.adsabs.harvard.edu/abs/2017MNRAS.470.2617A} {470, 2617}

\bibitem[\protect\citeauthoryear{{Alam}, {Peacock}, {Kraljic}, {Ross}  \& {Comparat}}{{Alam} et~al.}{2020}]{2020MNRAS.497..581A}
{Alam} S.,  {Peacock} J.~A.,  {Kraljic} K.,  {Ross} A.~J.,   {Comparat} J.,  2020, \mn@doi [\mnras] {10.1093/mnras/staa1956}, \href {https://ui.adsabs.harvard.edu/abs/2020MNRAS.497..581A} {497, 581}

\bibitem[\protect\citeauthoryear{{Alam} et~al.,}{{Alam} et~al.}{2021a}]{2021PhRvD.103h3533A}
{Alam} S.,  et~al., 2021a, \mn@doi [\prd] {10.1103/PhysRevD.103.083533}, \href {https://ui.adsabs.harvard.edu/abs/2021PhRvD.103h3533A} {103, 083533}

\bibitem[\protect\citeauthoryear{{Alam} et~al.,}{{Alam} et~al.}{2021b}]{2021MNRAS.504.4667A}
{Alam} S.,  et~al., 2021b, \mn@doi [\mnras] {10.1093/mnras/stab1150}, \href {https://ui.adsabs.harvard.edu/abs/2021MNRAS.504.4667A} {504, 4667}

\bibitem[\protect\citeauthoryear{{Alam}, {Paranjape}  \& {Peacock}}{{Alam} et~al.}{2023}]{2023arXiv230501266A}
{Alam} S.,  {Paranjape} A.,   {Peacock} J.~A.,  2023, \mn@doi [arXiv e-prints] {10.48550/arXiv.2305.01266}, \href {https://ui.adsabs.harvard.edu/abs/2023arXiv230501266A} {p. arXiv:2305.01266}

\bibitem[\protect\citeauthoryear{{Alexander} et~al.,}{{Alexander} et~al.}{2023}]{2023AJ....165..124A}
{Alexander} D.~M.,  et~al., 2023, \mn@doi [\aj] {10.3847/1538-3881/acacfc}, \href {https://ui.adsabs.harvard.edu/abs/2023AJ....165..124A} {165, 124}

\bibitem[\protect\citeauthoryear{{Allende Prieto} et~al.,}{{Allende Prieto} et~al.}{2020}]{2020RNAAS...4..188A}
{Allende Prieto} C.,  et~al., 2020, \mn@doi [Research Notes of the American Astronomical Society] {10.3847/2515-5172/abc1dc}, \href {https://ui.adsabs.harvard.edu/abs/2020RNAAS...4..188A} {4, 188}

\bibitem[\protect\citeauthoryear{{Bailey et al.}}{{Bailey et al.}}{2023}]{redrock2023}
{Bailey et al.} 2023, in preparation

\bibitem[\protect\citeauthoryear{{Balaguera-Antol{\'{\i}}nez}, {Kitaura}, {Pellejero-Ib{\'a}{\~n}ez}, {Zhao}  \& {Abel}}{{Balaguera-Antol{\'{\i}}nez} et~al.}{2019}]{2019MNRAS.483L..58B}
{Balaguera-Antol{\'{\i}}nez} A.,  {Kitaura} F.-S.,  {Pellejero-Ib{\'a}{\~n}ez} M.,  {Zhao} C.,   {Abel} T.,  2019, \mn@doi [\mnras] {10.1093/mnrasl/sly220}, \href {http://adsabs.harvard.edu/abs/2019MNRAS.483L..58B} {483, L58}

\bibitem[\protect\citeauthoryear{{Balaguera-Antol{\'\i}nez} et~al.,}{{Balaguera-Antol{\'\i}nez} et~al.}{2020}]{2020MNRAS.491.2565B}
{Balaguera-Antol{\'\i}nez} A.,  et~al., 2020, \mn@doi [\mnras] {10.1093/mnras/stz3206}, \href {https://ui.adsabs.harvard.edu/abs/2020MNRAS.491.2565B} {491, 2565}

\bibitem[\protect\citeauthoryear{{Balaguera-Antol{\'\i}nez} et~al.,}{{Balaguera-Antol{\'\i}nez} et~al.}{2022}]{2022arXiv221110640B}
{Balaguera-Antol{\'\i}nez} A.,  et~al., 2022, \mn@doi [arXiv e-prints] {10.48550/arXiv.2211.10640}, \href {https://ui.adsabs.harvard.edu/abs/2022arXiv221110640B} {p. arXiv:2211.10640}

\bibitem[\protect\citeauthoryear{{Baumgarten} \& {Chuang}}{{Baumgarten} \& {Chuang}}{2018}]{2018MNRAS.480.2535B}
{Baumgarten} F.,  {Chuang} C.-H.,  2018, \mn@doi [\mnras] {10.1093/mnras/sty1971}, \href {https://ui.adsabs.harvard.edu/abs/2018MNRAS.480.2535B} {480, 2535}

\bibitem[\protect\citeauthoryear{{Benson}, {Cole}, {Frenk}, {Baugh}  \& {Lacey}}{{Benson} et~al.}{2000}]{Benson2000}
{Benson} A.~J.,  {Cole} S.,  {Frenk} C.~S.,  {Baugh} C.~M.,   {Lacey} C.~G.,  2000, \mn@doi [\mnras] {10.1046/j.1365-8711.2000.03101.x}, \href {http://adsabs.harvard.edu/abs/2000MNRAS.311..793B} {311, 793}

\bibitem[\protect\citeauthoryear{{Berlind} \& {Weinberg}}{{Berlind} \& {Weinberg}}{2002}]{Berlind2002}
{Berlind} A.~A.,  {Weinberg} D.~H.,  2002, \mn@doi [\apj] {10.1086/341469}, \href {http://adsabs.harvard.edu/abs/2002ApJ...575..587B} {575, 587}

\bibitem[\protect\citeauthoryear{{Bernardeau}, {Colombi}, {Gazta{\~n}aga}  \& {Scoccimarro}}{{Bernardeau} et~al.}{2002}]{2002PhR...367....1B}
{Bernardeau} F.,  {Colombi} S.,  {Gazta{\~n}aga} E.,   {Scoccimarro} R.,  2002, \mn@doi [\physrep] {10.1016/S0370-1573(02)00135-7}, \href {https://ui.adsabs.harvard.edu/abs/2002PhR...367....1B} {367, 1}

\bibitem[\protect\citeauthoryear{{Brodzeller} et~al.,}{{Brodzeller} et~al.}{2023}]{2023arXiv230510426B}
{Brodzeller} A.,  et~al., 2023, \mn@doi [arXiv e-prints] {10.48550/arXiv.2305.10426}, \href {https://ui.adsabs.harvard.edu/abs/2023arXiv230510426B} {p. arXiv:2305.10426}

\bibitem[\protect\citeauthoryear{{Brouwer} et~al.,}{{Brouwer} et~al.}{2018}]{2018MNRAS.481.5189B}
{Brouwer} M.~M.,  et~al., 2018, \mn@doi [\mnras] {10.1093/mnras/sty2589}, \href {https://ui.adsabs.harvard.edu/abs/2018MNRAS.481.5189B} {481, 5189}

\bibitem[\protect\citeauthoryear{{Buchner} et~al.,}{{Buchner} et~al.}{2014}]{2014A&A...564A.125B}
{Buchner} J.,  et~al., 2014, \mn@doi [\aap] {10.1051/0004-6361/201322971}, \href {https://ui.adsabs.harvard.edu/abs/2014A&A...564A.125B} {564, A125}

\bibitem[\protect\citeauthoryear{{Chaussidon} et~al.,}{{Chaussidon} et~al.}{2023}]{2023ApJ...944..107C}
{Chaussidon} E.,  et~al., 2023, \mn@doi [\apj] {10.3847/1538-4357/acb3c2}, \href {https://ui.adsabs.harvard.edu/abs/2023ApJ...944..107C} {944, 107}

\bibitem[\protect\citeauthoryear{{Chaves-Montero}, {Angulo}  \& {Contreras}}{{Chaves-Montero} et~al.}{2023}]{2023MNRAS.521..937C}
{Chaves-Montero} J.,  {Angulo} R.~E.,   {Contreras} S.,  2023, \mn@doi [\mnras] {10.1093/mnras/stad243}, \href {https://ui.adsabs.harvard.edu/abs/2023MNRAS.521..937C} {521, 937}

\bibitem[\protect\citeauthoryear{Chuang, Kitaura, Prada, Zhao  \& Yepes}{Chuang et~al.}{2015}]{Chuang:2014vfa}
Chuang C.-H.,  Kitaura F.-S.,  Prada F.,  Zhao C.,   Yepes G.,  2015, \mn@doi [\mnras] {10.1093/mnras/stu2301}, 446, 2621

\bibitem[\protect\citeauthoryear{Chuang et~al.}{Chuang et~al.}{2019}]{Chuang:2018ega}
Chuang C.-H.,  et~al., 2019, \mn@doi [Mon. Not. Roy. Astron. Soc.] {10.1093/mnras/stz1233}, 487, 48

\bibitem[\protect\citeauthoryear{{Chuang et al.}}{{Chuang et al.}}{2023}]{DESI_MOCK_CHALLENGE_I}
{Chuang et al.} 2023, in preparation

\bibitem[\protect\citeauthoryear{{Cooper} et~al.,}{{Cooper} et~al.}{2022}]{2022arXiv220808514C}
{Cooper} A.~P.,  et~al., 2022, arXiv e-prints, \href {https://ui.adsabs.harvard.edu/abs/2022arXiv220808514C} {p. arXiv:2208.08514}

\bibitem[\protect\citeauthoryear{{Cooray} \& {Sheth}}{{Cooray} \& {Sheth}}{2002}]{Cooray2002}
{Cooray} A.,  {Sheth} R.,  2002, \mn@doi [\physrep] {10.1016/S0370-1573(02)00276-4}, \href {http://adsabs.harvard.edu/abs/2002PhR...372....1C} {372, 1}

\bibitem[\protect\citeauthoryear{{DESI Collaboration} et~al.,}{{DESI Collaboration} et~al.}{2016a}]{2016arXiv161100036D}
{DESI Collaboration} et~al., 2016a, arXiv e-prints, \href {https://ui.adsabs.harvard.edu/\#abs/2016arXiv161100036D} {p. arXiv:1611.00036}

\bibitem[\protect\citeauthoryear{{DESI Collaboration} et~al.,}{{DESI Collaboration} et~al.}{2016b}]{2016arXiv161100037D}
{DESI Collaboration} et~al., 2016b, arXiv e-prints, \href {https://ui.adsabs.harvard.edu/abs/2016arXiv161100037D} {p. arXiv:1611.00037}

\bibitem[\protect\citeauthoryear{{DESI Collaboration} et~al.,}{{DESI Collaboration} et~al.}{2022}]{2022AJ....164..207A}
{DESI Collaboration} et~al., 2022, \mn@doi [\aj] {10.3847/1538-3881/ac882b}, \href {https://ui.adsabs.harvard.edu/abs/2022AJ....164..207A} {164, 207}

\bibitem[\protect\citeauthoryear{{DESI Collaboration} et~al.,}{{DESI Collaboration} et~al.}{2023a}]{2023arXiv230606307D}
{DESI Collaboration} et~al., 2023a, \mn@doi [arXiv e-prints] {10.48550/arXiv.2306.06307}, \href {https://ui.adsabs.harvard.edu/abs/2023arXiv230606307D} {p. arXiv:2306.06307}

\bibitem[\protect\citeauthoryear{{DESI Collaboration} et~al.,}{{DESI Collaboration} et~al.}{2023b}]{2023arXiv230606308D}
{DESI Collaboration} et~al., 2023b, \mn@doi [arXiv e-prints] {10.48550/arXiv.2306.06308}, \href {https://ui.adsabs.harvard.edu/abs/2023arXiv230606308D} {p. arXiv:2306.06308}

\bibitem[\protect\citeauthoryear{{Dav{\'e}}, {Angl{\'e}s-Alc{\'a}zar}, {Narayanan}, {Li}, {Rafieferantsoa}  \& {Appleby}}{{Dav{\'e}} et~al.}{2019}]{2019MNRAS.486.2827D}
{Dav{\'e}} R.,  {Angl{\'e}s-Alc{\'a}zar} D.,  {Narayanan} D.,  {Li} Q.,  {Rafieferantsoa} M.~H.,   {Appleby} S.,  2019, \mn@doi [\mnras] {10.1093/mnras/stz937}, \href {https://ui.adsabs.harvard.edu/abs/2019MNRAS.486.2827D} {486, 2827}

\bibitem[\protect\citeauthoryear{{Dey} et~al.,}{{Dey} et~al.}{2019}]{2019AJ....157..168D}
{Dey} A.,  et~al., 2019, \mn@doi [\aj] {10.3847/1538-3881/ab089d}, \href {https://ui.adsabs.harvard.edu/abs/2019AJ....157..168D} {157, 168}

\bibitem[\protect\citeauthoryear{{Dubois} et~al.,}{{Dubois} et~al.}{2014}]{dubois2014}
{Dubois} Y.,  et~al., 2014, \mn@doi [\mnras] {10.1093/mnras/stu1227}, \href {http://adsabs.harvard.edu/abs/2014MNRAS.444.1453D} {444, 1453}

\bibitem[\protect\citeauthoryear{{Feng}, {Chu}, {Seljak}  \& {McDonald}}{{Feng} et~al.}{2016}]{Feng2016}
{Feng} Y.,  {Chu} M.-Y.,  {Seljak} U.,   {McDonald} P.,  2016, \mn@doi [\mnras] {10.1093/mnras/stw2123}, \href {https://ui.adsabs.harvard.edu/abs/2016MNRAS.463.2273F} {463, 2273}

\bibitem[\protect\citeauthoryear{{Feroz} \& {Hobson}}{{Feroz} \& {Hobson}}{2008}]{2008MNRAS.384..449F}
{Feroz} F.,  {Hobson} M.~P.,  2008, \mn@doi [\mnras] {10.1111/j.1365-2966.2007.12353.x}, \href {https://ui.adsabs.harvard.edu/abs/2008MNRAS.384..449F} {384, 449}

\bibitem[\protect\citeauthoryear{{Feroz}, {Hobson}  \& {Bridges}}{{Feroz} et~al.}{2009}]{2009MNRAS.398.1601F}
{Feroz} F.,  {Hobson} M.~P.,   {Bridges} M.,  2009, \mn@doi [\mnras] {10.1111/j.1365-2966.2009.14548.x}, \href {https://ui.adsabs.harvard.edu/abs/2009MNRAS.398.1601F} {398, 1601}

\bibitem[\protect\citeauthoryear{{Feroz}, {Hobson}, {Cameron}  \& {Pettitt}}{{Feroz} et~al.}{2019}]{2019OJAp....2E..10F}
{Feroz} F.,  {Hobson} M.~P.,  {Cameron} E.,   {Pettitt} A.~N.,  2019, \mn@doi [The Open Journal of Astrophysics] {10.21105/astro.1306.2144}, \href {https://ui.adsabs.harvard.edu/abs/2019OJAp....2E..10F} {2, 10}

\bibitem[\protect\citeauthoryear{{Gonzalez-Perez}, {Lacey}, {Baugh}, {Lagos}, {Helly}, {Campbell}  \& {Mitchell}}{{Gonzalez-Perez} et~al.}{2014}]{2014MNRAS.439..264G}
{Gonzalez-Perez} V.,  {Lacey} C.~G.,  {Baugh} C.~M.,  {Lagos} C.~D.~P.,  {Helly} J.,  {Campbell} D.~J.~R.,   {Mitchell} P.~D.,  2014, \mn@doi [\mnras] {10.1093/mnras/stt2410}, \href {https://ui.adsabs.harvard.edu/abs/2014MNRAS.439..264G} {439, 264}

\bibitem[\protect\citeauthoryear{{Grove} et~al.,}{{Grove} et~al.}{2022}]{2022MNRAS.515.1854G}
{Grove} C.,  et~al., 2022, \mn@doi [\mnras] {10.1093/mnras/stac1947}, \href {https://ui.adsabs.harvard.edu/abs/2022MNRAS.515.1854G} {515, 1854}

\bibitem[\protect\citeauthoryear{Gubner}{Gubner}{2006}]{johngubnerstats}
Gubner J.~A.,  2006, Probability and Random Processes for Electrical and Computer Engineers.
Cambridge University Press

\bibitem[\protect\citeauthoryear{{Guo} et~al.,}{{Guo} et~al.}{2011}]{2011MNRAS.413..101G}
{Guo} Q.,  et~al., 2011, \mn@doi [\mnras] {10.1111/j.1365-2966.2010.18114.x}, \href {https://ui.adsabs.harvard.edu/abs/2011MNRAS.413..101G} {413, 101}

\bibitem[\protect\citeauthoryear{{Guy} et~al.,}{{Guy} et~al.}{2023}]{2023AJ....165..144G}
{Guy} J.,  et~al., 2023, \mn@doi [\aj] {10.3847/1538-3881/acb212}, \href {https://ui.adsabs.harvard.edu/abs/2023AJ....165..144G} {165, 144}

\bibitem[\protect\citeauthoryear{{Hahn} et~al.,}{{Hahn} et~al.}{2022}]{2022arXiv220808512H}
{Hahn} C.,  et~al., 2022, arXiv e-prints, \href {https://ui.adsabs.harvard.edu/abs/2022arXiv220808512H} {p. arXiv:2208.08512}

\bibitem[\protect\citeauthoryear{{Hand}, {Feng}, {Beutler}, {Li}, {Modi}, {Seljak}  \& {Slepian}}{{Hand} et~al.}{2018}]{Hand2018}
{Hand} N.,  {Feng} Y.,  {Beutler} F.,  {Li} Y.,  {Modi} C.,  {Seljak} U.,   {Slepian} Z.,  2018, \mn@doi [\aj] {10.3847/1538-3881/aadae0}, \href {https://ui.adsabs.harvard.edu/abs/2018AJ....156..160H} {156, 160}

\bibitem[\protect\citeauthoryear{{Harnois-D{\'e}raps} \& {van Waerbeke}}{{Harnois-D{\'e}raps} \& {van Waerbeke}}{2015}]{2015MNRAS.450.2857H}
{Harnois-D{\'e}raps} J.,  {van Waerbeke} L.,  2015, \mn@doi [\mnras] {10.1093/mnras/stv794}, \href {https://ui.adsabs.harvard.edu/abs/2015MNRAS.450.2857H} {450, 2857}

\bibitem[\protect\citeauthoryear{{Harnois-D{\'e}raps}, {Pen}, {Iliev}, {Merz}, {Emberson}  \& {Desjacques}}{{Harnois-D{\'e}raps} et~al.}{2013}]{2013MNRAS.436..540H}
{Harnois-D{\'e}raps} J.,  {Pen} U.-L.,  {Iliev} I.~T.,  {Merz} H.,  {Emberson} J.~D.,   {Desjacques} V.,  2013, \mn@doi [\mnras] {10.1093/mnras/stt1591}, \href {https://ui.adsabs.harvard.edu/abs/2013MNRAS.436..540H} {436, 540}

\bibitem[\protect\citeauthoryear{{Harnois-D{\'e}raps} et~al.,}{{Harnois-D{\'e}raps} et~al.}{2018}]{2018MNRAS.481.1337H}
{Harnois-D{\'e}raps} J.,  et~al., 2018, \mn@doi [\mnras] {10.1093/mnras/sty2319}, \href {https://ui.adsabs.harvard.edu/abs/2018MNRAS.481.1337H} {481, 1337}

\bibitem[\protect\citeauthoryear{{Harnois-D{\'e}raps}, {Hernandez-Aguayo}, {Cuesta-Lazaro}, {Arnold}, {Li}, {Davies}  \& {Cai}}{{Harnois-D{\'e}raps} et~al.}{2022}]{2022arXiv221105779H}
{Harnois-D{\'e}raps} J.,  {Hernandez-Aguayo} C.,  {Cuesta-Lazaro} C.,  {Arnold} C.,  {Li} B.,  {Davies} C.~T.,   {Cai} Y.-C.,  2022, arXiv e-prints, \href {https://ui.adsabs.harvard.edu/abs/2022arXiv221105779H} {p. arXiv:2211.05779}

\bibitem[\protect\citeauthoryear{{Hartlap}, {Simon}  \& {Schneider}}{{Hartlap} et~al.}{2007}]{2007A&A...464..399H}
{Hartlap} J.,  {Simon} P.,   {Schneider} P.,  2007, \mn@doi [\aap] {10.1051/0004-6361:20066170}, \href {https://ui.adsabs.harvard.edu/abs/2007A&A...464..399H} {464, 399}

\bibitem[\protect\citeauthoryear{{Hearin} et~al.,}{{Hearin} et~al.}{2017}]{halotools}
{Hearin} A.~P.,  et~al., 2017, \mn@doi [The Astronomical Journal] {10.3847/1538-3881/aa859f}, \href {https://ui.adsabs.harvard.edu/abs/2017AJ....154..190H} {154, 190}

\bibitem[\protect\citeauthoryear{{Hildebrandt} et~al.,}{{Hildebrandt} et~al.}{2017}]{2017MNRAS.465.1454H}
{Hildebrandt} H.,  et~al., 2017, \mn@doi [\mnras] {10.1093/mnras/stw2805}, \href {https://ui.adsabs.harvard.edu/abs/2017MNRAS.465.1454H} {465, 1454}

\bibitem[\protect\citeauthoryear{{Joudaki} et~al.,}{{Joudaki} et~al.}{2017}]{2017MNRAS.465.2033J}
{Joudaki} S.,  et~al., 2017, \mn@doi [\mnras] {10.1093/mnras/stw2665}, \href {https://ui.adsabs.harvard.edu/abs/2017MNRAS.465.2033J} {465, 2033}

\bibitem[\protect\citeauthoryear{{Kirkby et al.}}{{Kirkby et al.}}{2023}]{expcalc}
{Kirkby et al.} 2023, in preparation

\bibitem[\protect\citeauthoryear{Kitaura, Yepes  \& Prada}{Kitaura et~al.}{2013}]{10.1093/mnrasl/slt172}
Kitaura F.-S.,  Yepes G.,   Prada F.,  2013, \mn@doi [\mnras: Letters] {10.1093/mnrasl/slt172}, 439, L21

\bibitem[\protect\citeauthoryear{{Kravtsov}, {Berlind}, {Wechsler}, {Klypin}, {Gottl{\"o}ber}, {Allgood}  \& {Primack}}{{Kravtsov} et~al.}{2004}]{2004ApJ...609...35K}
{Kravtsov} A.~V.,  {Berlind} A.~A.,  {Wechsler} R.~H.,  {Klypin} A.~A.,  {Gottl{\"o}ber} S.,  {Allgood} B.~o.,   {Primack} J.~R.,  2004, \mn@doi [\apj] {10.1086/420959}, \href {https://ui.adsabs.harvard.edu/abs/2004ApJ...609...35K} {609, 35}

\bibitem[\protect\citeauthoryear{{Lan} et~al.,}{{Lan} et~al.}{2023}]{2023ApJ...943...68L}
{Lan} T.-W.,  et~al., 2023, \mn@doi [\apj] {10.3847/1538-4357/aca5fa}, \href {https://ui.adsabs.harvard.edu/abs/2023ApJ...943...68L} {943, 68}

\bibitem[\protect\citeauthoryear{{Levi} et~al.,}{{Levi} et~al.}{2013}]{2013arXiv1308.0847L}
{Levi} M.,  et~al., 2013, arXiv e-prints, \href {https://ui.adsabs.harvard.edu/abs/2013arXiv1308.0847L} {p. arXiv:1308.0847}

\bibitem[\protect\citeauthoryear{Maksimova, Garrison, Eisenstein, Hadzhiyska, Bose  \& Satterthwaite}{Maksimova et~al.}{2021}]{Maksimova:2021ynf}
Maksimova N.~A.,  Garrison L.~H.,  Eisenstein D.~J.,  Hadzhiyska B.,  Bose S.,   Satterthwaite T.~P.,  2021, \mn@doi [Mon. Not. Roy. Astron. Soc.] {10.1093/mnras/stab2484}, 508, 4017

\bibitem[\protect\citeauthoryear{{Martinet} et~al.,}{{Martinet} et~al.}{2018}]{2018MNRAS.474..712M}
{Martinet} N.,  et~al., 2018, \mn@doi [\mnras] {10.1093/mnras/stx2793}, \href {https://ui.adsabs.harvard.edu/abs/2018MNRAS.474..712M} {474, 712}

\bibitem[\protect\citeauthoryear{{McCarthy}, {Schaye}, {Bird}  \& {Le Brun}}{{McCarthy} et~al.}{2017}]{2017MNRAS.465.2936M}
{McCarthy} I.~G.,  {Schaye} J.,  {Bird} S.,   {Le Brun} A. M.~C.,  2017, \mn@doi [\mnras] {10.1093/mnras/stw2792}, \href {https://ui.adsabs.harvard.edu/abs/2017MNRAS.465.2936M} {465, 2936}

\bibitem[\protect\citeauthoryear{{Miller} et~al.,}{{Miller} et~al.}{2023}]{2023arXiv230606310M}
{Miller} T.~N.,  et~al., 2023, \mn@doi [arXiv e-prints] {10.48550/arXiv.2306.06310}, \href {https://ui.adsabs.harvard.edu/abs/2023arXiv230606310M} {p. arXiv:2306.06310}

\bibitem[\protect\citeauthoryear{{Mo}, {van den Bosch}  \& {White}}{{Mo} et~al.}{2010}]{2010gfe..book.....M}
{Mo} H.,  {van den Bosch} F.~C.,   {White} S.,  2010, {Galaxy Formation and Evolution}.
Cambridge University Press

\bibitem[\protect\citeauthoryear{{Myers} et~al.,}{{Myers} et~al.}{2023}]{2023AJ....165...50M}
{Myers} A.~D.,  et~al., 2023, \mn@doi [\aj] {10.3847/1538-3881/aca5f9}, \href {https://ui.adsabs.harvard.edu/abs/2023AJ....165...50M} {165, 50}

\bibitem[\protect\citeauthoryear{{Navarro}, {Frenk}  \& {White}}{{Navarro} et~al.}{1996}]{1996ApJ...462..563N}
{Navarro} J.~F.,  {Frenk} C.~S.,   {White} S. D.~M.,  1996, \mn@doi [\apj] {10.1086/177173}, \href {https://ui.adsabs.harvard.edu/abs/1996ApJ...462..563N} {462, 563}

\bibitem[\protect\citeauthoryear{{Paranjape} \& {Alam}}{{Paranjape} \& {Alam}}{2020}]{2020MNRAS.495.3233P}
{Paranjape} A.,  {Alam} S.,  2020, \mn@doi [\mnras] {10.1093/mnras/staa1379}, \href {https://ui.adsabs.harvard.edu/abs/2020MNRAS.495.3233P} {495, 3233}

\bibitem[\protect\citeauthoryear{{Peacock} \& {Smith}}{{Peacock} \& {Smith}}{2000}]{Peacock2000}
{Peacock} J.~A.,  {Smith} R.~E.,  2000, \mn@doi [\mnras] {10.1046/j.1365-8711.2000.03779.x}, \href {http://adsabs.harvard.edu/abs/2000MNRAS.318.1144P} {318, 1144}

\bibitem[\protect\citeauthoryear{{Peebles} \& {Hauser}}{{Peebles} \& {Hauser}}{1974}]{1974ApJS...28...19P}
{Peebles} P.~J.~E.,  {Hauser} M.~G.,  1974, \mn@doi [\apjs] {10.1086/190308}, \href {https://ui.adsabs.harvard.edu/abs/1974ApJS...28...19P} {28, 19}

\bibitem[\protect\citeauthoryear{{Pellejero-Iba{\~n}ez} et~al.,}{{Pellejero-Iba{\~n}ez} et~al.}{2020}]{2020MNRAS.493..586P}
{Pellejero-Iba{\~n}ez} M.,  et~al., 2020, \mn@doi [\mnras] {10.1093/mnras/staa270}, \href {https://ui.adsabs.harvard.edu/abs/2020MNRAS.493..586P} {493, 586}

\bibitem[\protect\citeauthoryear{{Percival}, {Friedrich}, {Sellentin}  \& {Heavens}}{{Percival} et~al.}{2022}]{2022MNRAS.510.3207P}
{Percival} W.~J.,  {Friedrich} O.,  {Sellentin} E.,   {Heavens} A.,  2022, \mn@doi [\mnras] {10.1093/mnras/stab3540}, \href {https://ui.adsabs.harvard.edu/abs/2022MNRAS.510.3207P} {510, 3207}

\bibitem[\protect\citeauthoryear{{Pillepich} et~al.,}{{Pillepich} et~al.}{2018}]{2018MNRAS.473.4077P}
{Pillepich} A.,  et~al., 2018, \mn@doi [\mnras] {10.1093/mnras/stx2656}, \href {https://ui.adsabs.harvard.edu/abs/2018MNRAS.473.4077P} {473, 4077}

\bibitem[\protect\citeauthoryear{{Raichoor} et~al.,}{{Raichoor} et~al.}{2020}]{2020RNAAS...4..180R}
{Raichoor} A.,  et~al., 2020, \mn@doi [Research Notes of the American Astronomical Society] {10.3847/2515-5172/abc078}, \href {https://ui.adsabs.harvard.edu/abs/2020RNAAS...4..180R} {4, 180}

\bibitem[\protect\citeauthoryear{{Raichoor et al.}}{{Raichoor et al.}}{2023a}]{fba}
{Raichoor et al.} 2023a, in preparation

\bibitem[\protect\citeauthoryear{{Raichoor} et~al.,}{{Raichoor} et~al.}{2023b}]{2023AJ....165..126R}
{Raichoor} A.,  et~al., 2023b, \mn@doi [\aj] {10.3847/1538-3881/acb213}, \href {https://ui.adsabs.harvard.edu/abs/2023AJ....165..126R} {165, 126}

\bibitem[\protect\citeauthoryear{{Ruiz-Macias} et~al.,}{{Ruiz-Macias} et~al.}{2020}]{2020RNAAS...4..187R}
{Ruiz-Macias} O.,  et~al., 2020, \mn@doi [Research Notes of the American Astronomical Society] {10.3847/2515-5172/abc25a}, \href {https://ui.adsabs.harvard.edu/abs/2020RNAAS...4..187R} {4, 187}

\bibitem[\protect\citeauthoryear{{Schaye} et~al.,}{{Schaye} et~al.}{2010}]{2010MNRAS.402.1536S}
{Schaye} J.,  et~al., 2010, \mn@doi [\mnras] {10.1111/j.1365-2966.2009.16029.x}, \href {https://ui.adsabs.harvard.edu/\#abs/2010MNRAS.402.1536S} {402, 1536}

\bibitem[\protect\citeauthoryear{{Schaye} et~al.,}{{Schaye} et~al.}{2015}]{2015MNRAS.446..521S}
{Schaye} J.,  et~al., 2015, \mn@doi [\mnras] {10.1093/mnras/stu2058}, \href {https://ui.adsabs.harvard.edu/abs/2015MNRAS.446..521S} {446, 521}

\bibitem[\protect\citeauthoryear{{Schlafly} et~al.,}{{Schlafly} et~al.}{2023}]{2023arXiv230606309S}
{Schlafly} E.~F.,  et~al., 2023, \mn@doi [arXiv e-prints] {10.48550/arXiv.2306.06309}, \href {https://ui.adsabs.harvard.edu/abs/2023arXiv230606309S} {p. arXiv:2306.06309}

\bibitem[\protect\citeauthoryear{{Schlegel et al.}}{{Schlegel et al.}}{2023}]{dr9}
{Schlegel et al.} 2023, in preparation

\bibitem[\protect\citeauthoryear{Sefusatti, Crocce, Scoccimarro  \& Couchman}{Sefusatti et~al.}{2016}]{10.1093/mnras/stw1229}
Sefusatti E.,  Crocce M.,  Scoccimarro R.,   Couchman H. M.~P.,  2016, \mn@doi [\mnras] {10.1093/mnras/stw1229}, 460, 3624

\bibitem[\protect\citeauthoryear{{Seljak}}{{Seljak}}{2000}]{Seljak2000}
{Seljak} U.,  2000, \mn@doi [\mnras] {10.1046/j.1365-8711.2000.03715.x}, \href {http://adsabs.harvard.edu/abs/2000MNRAS.318..203S} {318, 203}

\bibitem[\protect\citeauthoryear{{Sellentin} \& {Heavens}}{{Sellentin} \& {Heavens}}{2016}]{2016MNRAS.456L.132S}
{Sellentin} E.,  {Heavens} A.~F.,  2016, \mn@doi [\mnras] {10.1093/mnrasl/slv190}, \href {https://ui.adsabs.harvard.edu/abs/2016MNRAS.456L.132S} {456, L132}

\bibitem[\protect\citeauthoryear{{Sheth}, {Mo}  \& {Tormen}}{{Sheth} et~al.}{2001}]{2001MNRAS.323....1S}
{Sheth} R.~K.,  {Mo} H.~J.,   {Tormen} G.,  2001, \mn@doi [\mnras] {10.1046/j.1365-8711.2001.04006.x}, \href {https://ui.adsabs.harvard.edu/abs/2001MNRAS.323....1S} {323, 1}

\bibitem[\protect\citeauthoryear{{Silber} et~al.,}{{Silber} et~al.}{2023}]{2023AJ....165....9S}
{Silber} J.~H.,  et~al., 2023, \mn@doi [\aj] {10.3847/1538-3881/ac9ab1}, \href {https://ui.adsabs.harvard.edu/abs/2023AJ....165....9S} {165, 9}

\bibitem[\protect\citeauthoryear{{Tamone} et~al.,}{{Tamone} et~al.}{2020}]{2020MNRAS.499.5527T}
{Tamone} A.,  et~al., 2020, \mn@doi [\mnras] {10.1093/mnras/staa3050}, \href {https://ui.adsabs.harvard.edu/abs/2020MNRAS.499.5527T} {499, 5527}

\bibitem[\protect\citeauthoryear{{Tasitsiomi}, {Kravtsov}, {Wechsler}  \& {Primack}}{{Tasitsiomi} et~al.}{2004}]{2004ApJ...614..533T}
{Tasitsiomi} A.,  {Kravtsov} A.~V.,  {Wechsler} R.~H.,   {Primack} J.~R.,  2004, \mn@doi [\apj] {10.1086/423784}, \href {https://ui.adsabs.harvard.edu/abs/2004ApJ...614..533T} {614, 533}

\bibitem[\protect\citeauthoryear{{Vale} \& {Ostriker}}{{Vale} \& {Ostriker}}{2004}]{2004MNRAS.353..189V}
{Vale} A.,  {Ostriker} J.~P.,  2004, \mn@doi [\mnras] {10.1111/j.1365-2966.2004.08059.x}, \href {https://ui.adsabs.harvard.edu/abs/2004MNRAS.353..189V} {353, 189}

\bibitem[\protect\citeauthoryear{{Wadekar} \& {Scoccimarro}}{{Wadekar} \& {Scoccimarro}}{2020}]{Wad20a}
{Wadekar} D.,  {Scoccimarro} R.,  2020, \mn@doi [\prd] {10.1103/PhysRevD.102.123517}, \href {https://ui.adsabs.harvard.edu/abs/2020PhRvD.102l3517W} {102, 123517}

\bibitem[\protect\citeauthoryear{{Wadekar}, {Ivanov}  \& {Scoccimarro}}{{Wadekar} et~al.}{2020}]{Wad20b}
{Wadekar} D.,  {Ivanov} M.~M.,   {Scoccimarro} R.,  2020, \mn@doi [\prd] {10.1103/PhysRevD.102.123521}, \href {https://ui.adsabs.harvard.edu/abs/2020PhRvD.102l3521W} {102, 123521}

\bibitem[\protect\citeauthoryear{{Wechsler} \& {Tinker}}{{Wechsler} \& {Tinker}}{2018}]{2018ARA&A..56..435W}
{Wechsler} R.~H.,  {Tinker} J.~L.,  2018, \mn@doi [\araa] {10.1146/annurev-astro-081817-051756}, \href {https://ui.adsabs.harvard.edu/abs/2018ARA&A..56..435W} {56, 435}

\bibitem[\protect\citeauthoryear{{White}, {Hernquist}  \& {Springel}}{{White} et~al.}{2001}]{White2001}
{White} M.,  {Hernquist} L.,   {Springel} V.,  2001, \mn@doi [\apjl] {10.1086/319644}, \href {http://adsabs.harvard.edu/abs/2001ApJ...550L.129W} {550, L129}

\bibitem[\protect\citeauthoryear{{Xu}, {Cuesta}, {Padmanabhan}, {Eisenstein}  \& {McBride}}{{Xu} et~al.}{2013}]{Xu13}
{Xu} X.,  {Cuesta} A.~J.,  {Padmanabhan} N.,  {Eisenstein} D.~J.,   {McBride} C.~K.,  2013, \mn@doi [\mnras] {10.1093/mnras/stt379}, \href {http://adsabs.harvard.edu/abs/2013MNRAS.431.2834X} {431, 2834}

\bibitem[\protect\citeauthoryear{{Y{\`e}che} et~al.,}{{Y{\`e}che} et~al.}{2020}]{2020RNAAS...4..179Y}
{Y{\`e}che} C.,  et~al., 2020, \mn@doi [Research Notes of the American Astronomical Society] {10.3847/2515-5172/abc01a}, \href {https://ui.adsabs.harvard.edu/abs/2020RNAAS...4..179Y} {4, 179}

\bibitem[\protect\citeauthoryear{Zarrouk et~al.}{Zarrouk et~al.}{2021}]{Zarrouk:2020hha}
Zarrouk P.,  et~al., 2021, \mn@doi [Mon. Not. Roy. Astron. Soc.] {10.1093/mnras/stab298}, 503, 2562

\bibitem[\protect\citeauthoryear{{Zel'dovich}}{{Zel'dovich}}{1970}]{1970A&A.....5...84Z}
{Zel'dovich} Y.~B.,  1970, \aap, \href {https://ui.adsabs.harvard.edu/abs/1970A&A.....5...84Z} {5, 84}

\bibitem[\protect\citeauthoryear{{Zhang et al.}}{{Zhang et al.}}{2023}]{Zhang_jackknife_paper}
{Zhang et al.} 2023, in preparation

\bibitem[\protect\citeauthoryear{{Zhao}}{{Zhao}}{2023}]{2023arXiv230112557Z}
{Zhao} C.,  2023, \mn@doi [arXiv e-prints] {10.48550/arXiv.2301.12557}, \href {https://ui.adsabs.harvard.edu/abs/2023arXiv230112557Z} {p. arXiv:2301.12557}

\bibitem[\protect\citeauthoryear{{Zhao} et~al.,}{{Zhao} et~al.}{2021}]{2021MNRAS.503.1149Z}
{Zhao} C.,  et~al., 2021, \mn@doi [\mnras] {10.1093/mnras/stab510}, \href {https://ui.adsabs.harvard.edu/abs/2021MNRAS.503.1149Z} {503, 1149}

\bibitem[\protect\citeauthoryear{Zheng et~al.,}{Zheng et~al.}{2005}]{Zheng_2005}
Zheng Z.,  et~al., 2005, \mn@doi [The Astrophysical Journal] {10.1086/466510}, 633, 791

\bibitem[\protect\citeauthoryear{{Zhou} et~al.,}{{Zhou} et~al.}{2020}]{2020RNAAS...4..181Z}
{Zhou} R.,  et~al., 2020, \mn@doi [Research Notes of the American Astronomical Society] {10.3847/2515-5172/abc0f4}, \href {https://ui.adsabs.harvard.edu/abs/2020RNAAS...4..181Z} {4, 181}

\bibitem[\protect\citeauthoryear{{Zhou} et~al.,}{{Zhou} et~al.}{2023}]{2023AJ....165...58Z}
{Zhou} R.,  et~al., 2023, \mn@doi [\aj] {10.3847/1538-3881/aca5fb}, \href {https://ui.adsabs.harvard.edu/abs/2023AJ....165...58Z} {165, 58}

\bibitem[\protect\citeauthoryear{{Zou} et~al.,}{{Zou} et~al.}{2017}]{2017PASP..129f4101Z}
{Zou} H.,  et~al., 2017, \mn@doi [\pasp] {10.1088/1538-3873/aa65ba}, \href {https://ui.adsabs.harvard.edu/abs/2017PASP..129f4101Z} {129, 064101}

\bibitem[\protect\citeauthoryear{{de Mattia} et~al.,}{{de Mattia} et~al.}{2021}]{2021MNRAS.501.5616D}
{de Mattia} A.,  et~al., 2021, \mn@doi [\mnras] {10.1093/mnras/staa3891}, \href {https://ui.adsabs.harvard.edu/abs/2021MNRAS.501.5616D} {501, 5616}

\bibitem[\protect\citeauthoryear{{van Uitert} et~al.,}{{van Uitert} et~al.}{2018}]{2018MNRAS.476.4662V}
{van Uitert} E.,  et~al., 2018, \mn@doi [\mnras] {10.1093/mnras/sty551}, \href {https://ui.adsabs.harvard.edu/abs/2018MNRAS.476.4662V} {476, 4662}

\makeatother
\end{thebibliography}



\section*{Affiliations}
\small $^{1}$Institute of Physics, Laboratory of Astrophysics, École Polytechnique Fédérale de Lausanne (EPFL), Observatoire de Sauverny, CH-1290 Versoix, Switzerland\\
\small $^{2}$Tata Institute of Fundamental Research, Homi Bhabha Road, Mumbai 400005, India\\
\small $^{3}$Institute for Astronomy, University of Edinburgh, Royal Observatory, Blackford Hill, Edinburgh EH9 3HJ, UK\\
\small $^{4}$Department of Physics and Astronomy, The University of Utah, 115 South 1400 East, Salt Lake City, UT 84112, USA\\
\small $^{5}$Kavli Institute for Particle Astrophysics and Cosmology, Stanford University, 452 Lomita Mall, Stanford, CA 94305, USA\\
\small $^{6}$Department of Astronomy, School of Physics and Astronomy, Shanghai Jiao Tong University, Shanghai 200240, China\\
\small $^{7}$Key Laboratory for Particle Astrophysics and Cosmology(MOE)/Shanghai Key Laboratory for Particle Physics and Cosmology, China\\
\small $^{8}$Lawrence Berkeley National Laboratory, 1 Cyclotron Road, Berkeley, CA 94720, USA\\
\small $^{9}$Physics Dept., Boston University, 590 Commonwealth Avenue, Boston, MA 02215, USA\\
\small $^{10}$Department of Physics \& Astronomy, University College London, Gower Street, London, WC1E 6BT, UK\\
\small $^{11}$Institute for Computational Cosmology, Department of Physics, Durham University, South Road, Durham DH1 3LE, UK\\
\small $^{12}$Instituto de F\'{\i}sica, Universidad Nacional Aut\'{o}noma de M\'{e}xico,  Cd. de M\'{e}xico  C.P. 04510,  M\'{e}xico\\
\small $^{13}$Departamento de F\'isica, Universidad de los Andes, Cra. 1 No. 18A-10, Edificio Ip, CP 111711, Bogot\'a, Colombia\\
\small $^{14}$Observatorio Astron\'omico, Universidad de los Andes, Cra. 1 No. 18A-10, Edificio H, CP 111711 Bogot\'a, Colombia\\
\small $^{15}$Center for Cosmology and AstroParticle Physics, The Ohio State University, 191 West Woodruff Avenue, Columbus, OH 43210, USA\\
\small $^{16}$Department of Physics, The Ohio State University, 191 West Woodruff Avenue, Columbus, OH 43210, USA\\
\small $^{17}$The Ohio State University, Columbus, 43210 OH, USA\\
\small $^{18}$Departament de F\'{i}sica, Serra H\'{u}nter, Universitat Aut\`{o}noma de Barcelona, 08193 Bellaterra (Barcelona), Spain\\
\small $^{19}$Institut de F\'{i}sica d’Altes Energies (IFAE), The Barcelona Institute of Science and Technology, Campus UAB, 08193 Bellaterra Barcelona, Spain\\
\small $^{20}$Instituci\'{o} Catalana de Recerca i Estudis Avan\c{c}ats, Passeig de Llu\'{\i}s Companys, 23, 08010 Barcelona, Spain\\
\small $^{21}$National Astronomical Observatories, Chinese Academy of Sciences, A20 Datun Rd., Chaoyang District, Beijing, 100012, P.R. China\\
\small $^{22}$Department of Physics and Astronomy, University of Waterloo, 200 University Ave W, Waterloo, ON N2L 3G1, Canada\\
\small $^{23}$Perimeter Institute for Theoretical Physics, 31 Caroline St. North, Waterloo, ON N2L 2Y5, Canada\\
\small $^{24}$Waterloo Centre for Astrophysics, University of Waterloo, 200 University Ave W, Waterloo, ON N2L 3G1, Canada\\
\small $^{25}$Space Sciences Laboratory, University of California, Berkeley, 7 Gauss Way, Berkeley, CA  94720, USA\\
\small $^{26}$University of California, Berkeley, 110 Sproul Hall \#5800 Berkeley, CA 94720, USA\\
\small $^{27}$Department of Physics, Kansas State University, 116 Cardwell Hall, Manhattan, KS 66506, USA\\
\small $^{28}$Department of Physics and Astronomy, Sejong University, Seoul, 143-747, Korea\\
\small $^{29}$CIEMAT, Avenida Complutense 40, E-28040 Madrid, Spain\\
\small $^{30}$Department of Physics, University of Michigan, Ann Arbor, MI 48109, USA\\
\small $^{31}$University of Michigan, Ann Arbor, MI 48109, USA\\
\small $^{32}$Department of Physics \& Astronomy, Ohio University, Athens, OH 45701, USA\\


\appendix

\section{Convergence tests}
\label{sec:appendix_robustnesstest}

Given the two-step approach to fit the HOD model, there is the important question of convergence that has to be answered. Thus, we study whether the $\Sigma_\mathrm{diff}$ estimates well the noise in our HOD fitting process.

\begin{figure*}
	\includegraphics[width=\textwidth]{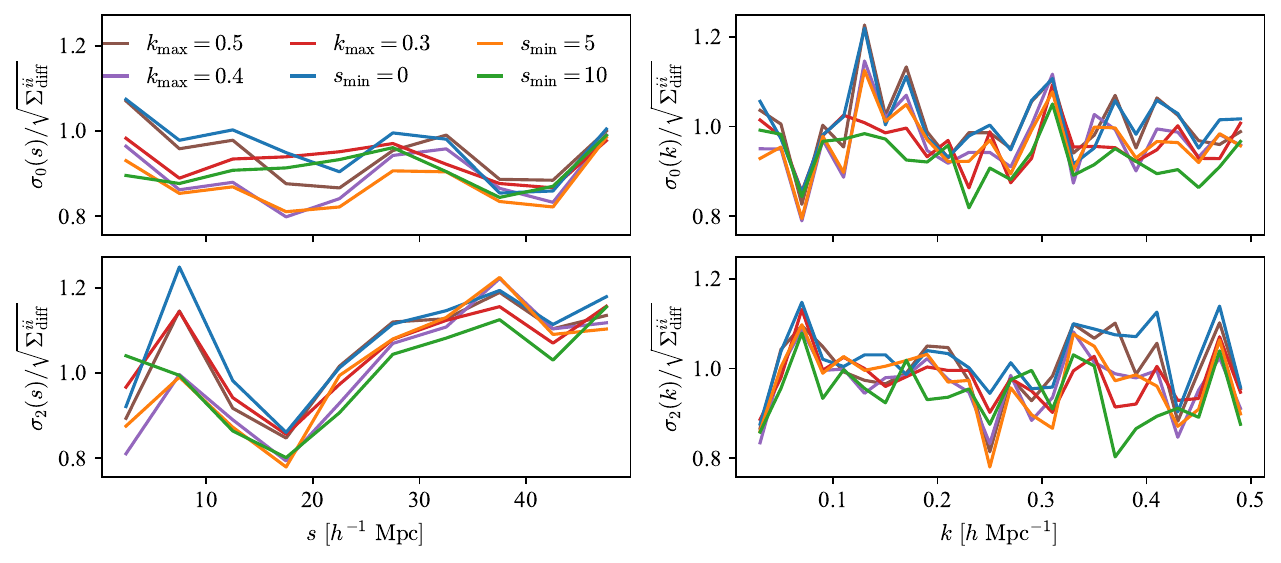}
    \caption{The ratios between the standard deviations computed on the differences of $N_\mathrm{mocks}^\mathrm{cov}=123$ (LR \textsc{FastPM}, \textsc{SLICS}) clustering pairs and the square root of the diagonal of $\Sigma_\mathrm{diff}$, i.e. $\Sigma^{ii}_\mathrm{diff}$. The colours denote the HOD fitting scenarios in the second HOD fitting step, see Table~\ref{tab:fitting_intervals_n_bins}. While the left panels include monopole and quadrupole of the 2PCF, the right ones display the power spectrum.}
    \label{fig:ratio_clustering_error_bars}
\end{figure*}

Figure~\ref{fig:ratio_clustering_error_bars} illustrates that the magnitude of the errors estimated after the second HOD fitting step for all six hod fitting scenarios are consistent between themselves within at most 10 to 20 per cent. However, there seems to be a slight divergence between the error estimations when one approaches the lowest scales.

\begin{figure*}
	\includegraphics[width=\textwidth]{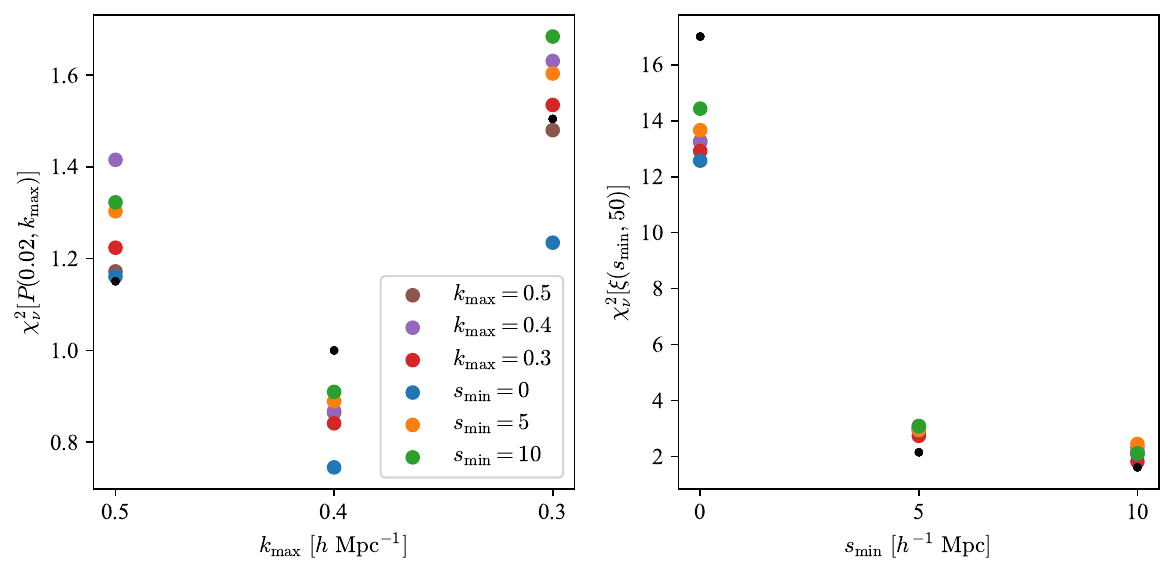}
    \caption{The $\chi_\nu^2$ as defined in Section~\ref{sec:reduced_chi2}, but using different covariance matrices. We compute $\chi_\nu^2$: 1) for the six best-fitting \textsc{FastPM} cases, three cases for the power spectrum in the left panel (see Section~\ref{sec:power_spectrum_fitting}), and three cases for the 2PCF in the right panel (see Section~\ref{sec:2pcf_fitting}); 2) with the six difference covariance matrices obtained after the second HOD fitting step (the coloured dots). The black points show the best fitting $\chi_\nu^2$ for the six cases that also appear in Figure~\ref{fig:reduced_chi2}. }
    \label{fig:chi2_after_second_step_6_covariances}
\end{figure*}

Finally, compared to the square root of the diagonal of $\Sigma_\mathrm{diff}$, the six standard deviations for both 2PCF and power spectrum are found within at most 20 per cent deviation. In order to quantify these discrepancies, we have built the difference covariance matrix using each of the six best-fitting \textsc{FastPM} (of the second HOD fitting step, see Table~\ref{tab:fitting_intervals_n_bins}) set of galaxies. Furthermore, we have computed the $\chi_\nu^2$ for the same best-fitting \textsc{FastPM} of the second HOD fitting step, but using these six new difference covariance matrices. The results are summarised in Figure~\ref{fig:chi2_after_second_step_6_covariances}.

Even though for most of the fitting cases, the six new difference covariance matrices seem to provide coherent biased $\chi^2_\nu$ values compared to the $\Sigma_\mathrm{diff}$, the biases do not share the same sign between the fitting cases. In addition, most $\chi^2_\nu$ values are within the error bars shown in Table~\ref{tab:error_on_chisq} with respect to the reference. This suggests that given the error bars, the hypothetical best-fittings obtained using the six new covariance matrices, would be indistinguishable from the best-fitting \textsc{FastPM} of the second HOD step. Consequently, we argue that the $\Sigma_\mathrm{diff}$ is a good approximation of the noise in the difference of the (\textsc{FastPM}, \textsc{SLICS}) clustering, thus a hypothetical third step HOD would not drastically change the best-fitting \textsc{FastPM} compared to the ones after the second step. 

\section{Uncertainty of the goodness-of-fit}
\label{sec:appendix_chi2uncertainty}

In this section, we are studying the uncertainty introduced by the covariance matrix and the finite number of HOD realisations per \textsc{FastPM} halo catalogue in the values of $\chi^2_\nu$, as defined in Eq.~\eqref{eq:reduced_chi2}. The results are summarised in Table~\ref{tab:error_on_chisq}.

\subsection{Covariance matrix induced uncertainty}
Due to the fact that we have only $N^\mathrm{cov}_\mathrm{mocks} = 123$ \textsc{SLICS} and \textsc{FastPM} realisations that share the same initial conditions, we are bound to use the JackKnife (JK) method to estimate the uncertainty introduced by the covariance matrix. Additionally, the HOD fitting is computationally expensive ($\approx6000$ CPU-hours), thus we are not able to perform hundreds of HOD fittings with different covariance matrices. Consequently, after obtaining one set of best-fitting HOD parameters, we computed the $\chi^2_\nu$ with the same best-fitting \textsc{FastPM} clustering, but with $N^\mathrm{cov}_\mathrm{mocks}$ different covariance matrices. 

The covariance matrices -- $\Sigma_\chi^i$, with $i$ from 1 to $N^\mathrm{cov}_\mathrm{mocks}$ -- are estimated using Eq.~\eqref{eq:inverse_unbiased_covaraince}, but with only $N^\mathrm{cov}_\mathrm{mocks}-1$ clustering realisations.
Furthermore, we compute $\chi^{2,i}_{\nu, \mathrm{JK}}$ for each $\Sigma_\chi^i$, as defined in Eq.~\eqref{eq:reduced_chi2} and we calculate the mean $\bar{\chi}^2_{\nu,\mathrm{JK}}$ and the variance $\sigma_{\chi,\mathrm{JK}}^2$:
\begin{equation}
    \label{eq:appendix_chi2_jk}
    \bar{\chi}^2_{\nu,\mathrm{JK}} = \frac{1}{N^\mathrm{cov}_\mathrm{mocks}}\sum_{i=1}^{N^\mathrm{cov}_\mathrm{mocks}} \chi^{2,i}_{\nu, \mathrm{JK}},
\end{equation}
\begin{equation}
    \label{eq:appendix_chi2_var_jk}
    \sigma_{\chi,\mathrm{JK}}^2 = \left[ \frac{N^\mathrm{cov}_\mathrm{mocks}-1}{N^\mathrm{cov}_\mathrm{mocks}} \sum_{i=1}^{N^\mathrm{cov}_\mathrm{mocks}} \left(\chi^{2,i}_{\nu, \mathrm{JK}} - \bar{\chi}^2_{\nu,\mathrm{JK}} \right)^2 \right].
\end{equation}

\subsection{HOD induced uncertainty}
\begin{table}
    \centering
    \begin{tabular}{l | l | l }
        \hline
                & $P(k)$ & $\xi(s)$   \\
         \hline
        Large                                                            &               &                \\
        $\chi^2_\nu$ from Figure~\ref{fig:reduced_chi2}                  & $1.15$        & $16.94$        \\
        $\bar{\chi}^2_{\nu,\mathrm{HOD}} \pm \sigma_{\chi,\mathrm{HOD}}$ & $1.38\pm0.26$ & $17.03\pm1.65$ \\
        $\bar{\chi}^2_{\nu,\mathrm{JK}}  \pm \sigma_{\chi,\mathrm{JK}}$  & $1.16\pm0.31$ & $16.97\pm2.68$ \\
                                                                    \\           
         Medium &                        &                          \\
                & $1.00$                 & $2.16$                   \\
                & $1.13\pm0.23$          & $2.19\pm0.42$            \\
                & $1.00\pm0.19$          & $2.16\pm0.32$            \\
                                                                    \\
         Small  &                        &                          \\
                & $1.50$                 & $1.59$                   \\
                & $1.50\pm0.29$          & $1.68\pm0.41$            \\
                & $1.51\pm0.25$          & $1.59\pm0.26$            \\
         \hline
         
    \end{tabular}
    \caption{The values of the $\chi^2_\nu$ and their uncertainties introduced by the covariance matrix (Eqs.~\eqref{eq:appendix_chi2_jk} and \eqref{eq:appendix_chi2_var_jk}) and the finite number of HOD realisations per \textsc{FastPM} halo catalogue (Eqs.~\eqref{eq:appendix_chi2_hod} and \eqref{eq:appendix_chi2_var_hod}). The estimations have been performed on the LR ($1296^3$) \textsc{FastPM} galaxy catalogues and on both the 2PCF and the power spectrum for the three specific fitting ranges defined in Table~\ref{tab:fitting_intervals_n_bins}. \label{tab:error_on_chisq}} 
\end{table}

During the HOD fitting, for each \textsc{FastPM} halo catalogue we create a single galaxy realisation, in order to reduce the optimisation time. As a consequence, we introduce additional noise in the HOD fitting process, that is not considered in the covariance matrix.

With the aim of estimating the effect of this noise on the $\chi^2_\nu$, we compute 100 galaxy realisations for a given set of best-fitting HOD parameters and per \textsc{FastPM} halo catalogue. Furthermore, using the 20 galaxy realisations corresponding to the halo catalogues used in the HOD fitting process and the same covariance matrix, we compute $\chi^{2,i}_{\nu, \mathrm{HOD}}$ as in Eq.~\eqref{eq:reduced_chi2}, where $i=1,...,100$. Finally, we calculate the mean and the standard deviation of the 100 $\chi^{2,i}_{\nu, \mathrm{HOD}}$ values:
\begin{equation}
    \label{eq:appendix_chi2_hod}
    \bar{\chi}^2_{\nu,\mathrm{HOD}} = \frac{1}{100}\sum_{i=1}^{100} \chi^{2,i}_{\nu, \mathrm{HOD}},
\end{equation}
\begin{equation}
    \label{eq:appendix_chi2_var_hod}
    \sigma_{\chi,\mathrm{HOD}}^2 = \left[ \frac{1}{100 - 1} \sum_{i=1}^{100} \left(\chi^{2,i}_{\nu, \mathrm{HOD}} - \bar{\chi}^2_{\nu,\mathrm{HOD}} \right)^2 \right].
\end{equation}

\section{Covariance matrix comparison}
\label{sec:appendix_cov_mat_comp}

\begin{figure}
	\includegraphics[width=\columnwidth]{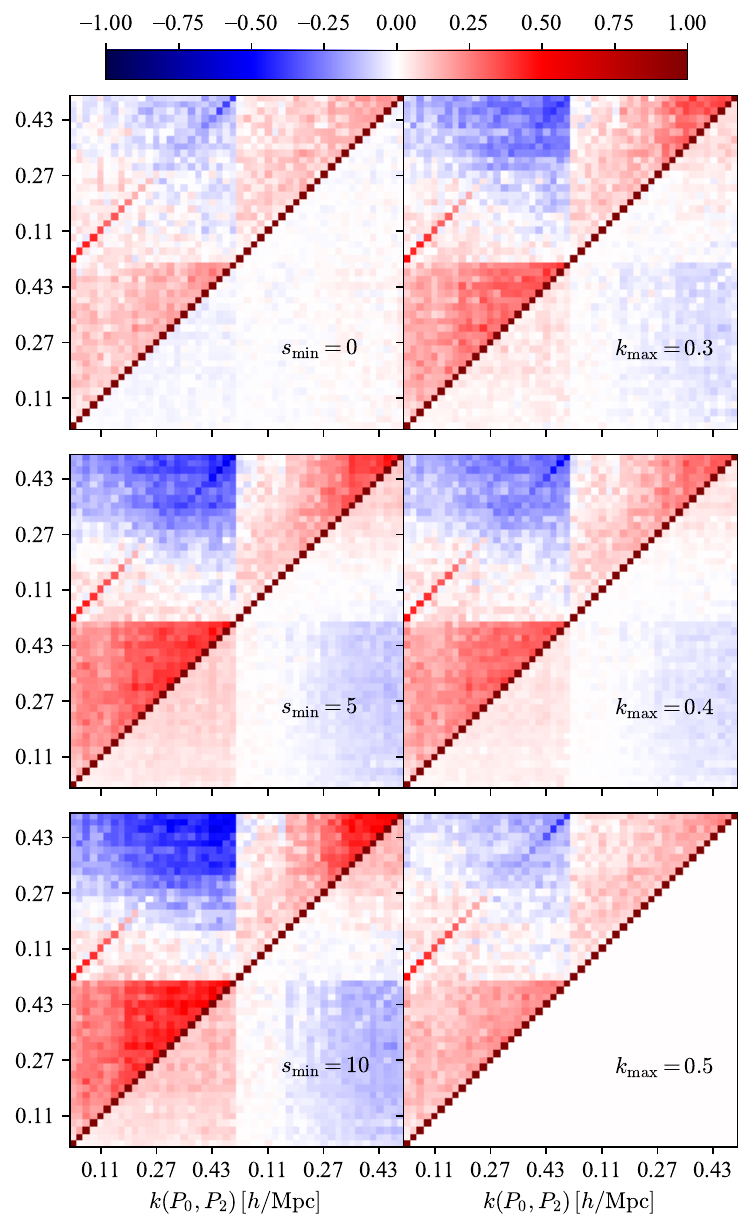}
    \caption{Upper triangular matrices: correlation matrices of the power spectrum monopole and quadrupole, for the fitting cases defined in Table~\ref{tab:fitting_intervals_n_bins}. Lower triangular matrices: the difference between the shown correlation matrix and the reference one, i.e. $k_\mathrm{max}=0.5$.}
    \label{fig:correlation_pspec_1296_780}
\end{figure}

\begin{figure}
	\includegraphics[width=\columnwidth]{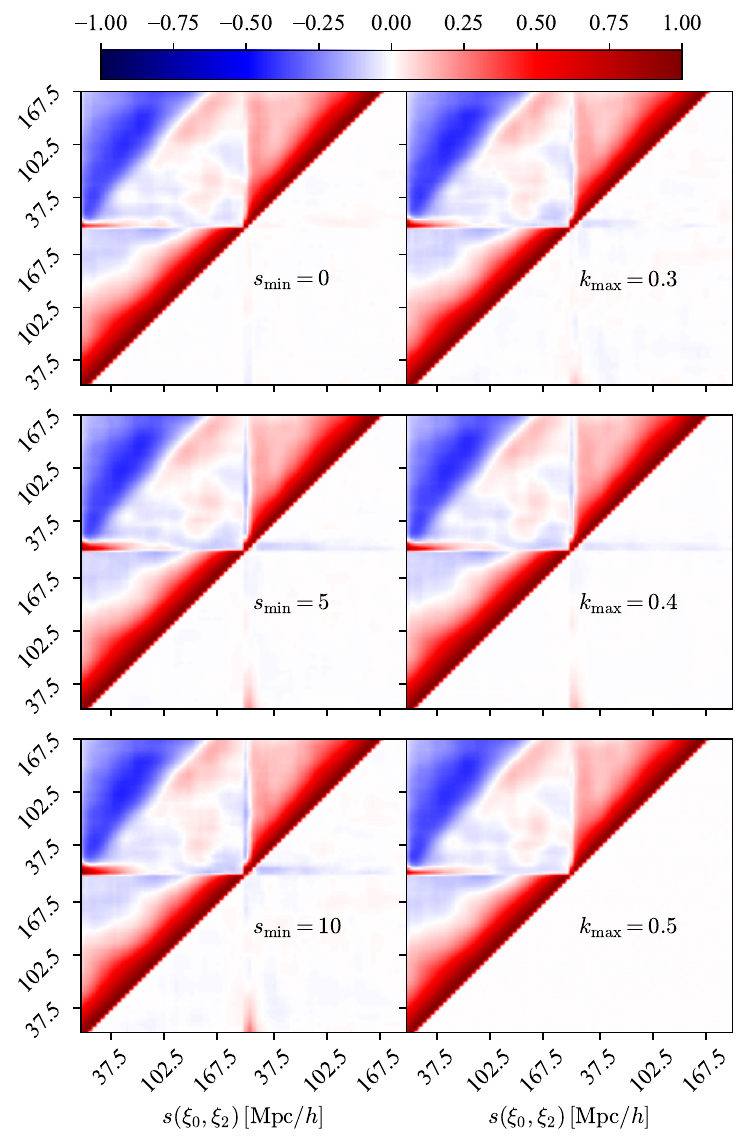}
    \caption{Same as Figure~\ref{fig:correlation_pspec_1296_780}, but for 2PCF}
    \label{fig:correlation_2pcf_1296_780}
\end{figure}

Analysing Figures~\ref{fig:correlation_pspec_1296_780} and \ref{fig:correlation_2pcf_1296_780} one can observe that the ($s_\mathrm{min}=5$, $s_\mathrm{min}=10$) and ($k_\mathrm{max}=0.3$, $k_\mathrm{max}=0.4$) pairs have very similar correlation matrices. Consequently, we only show $k_\mathrm{max}=0.3$ and $s_\mathrm{min}=10$ in the main text.

\begin{figure*}
	\includegraphics[width=\textwidth]{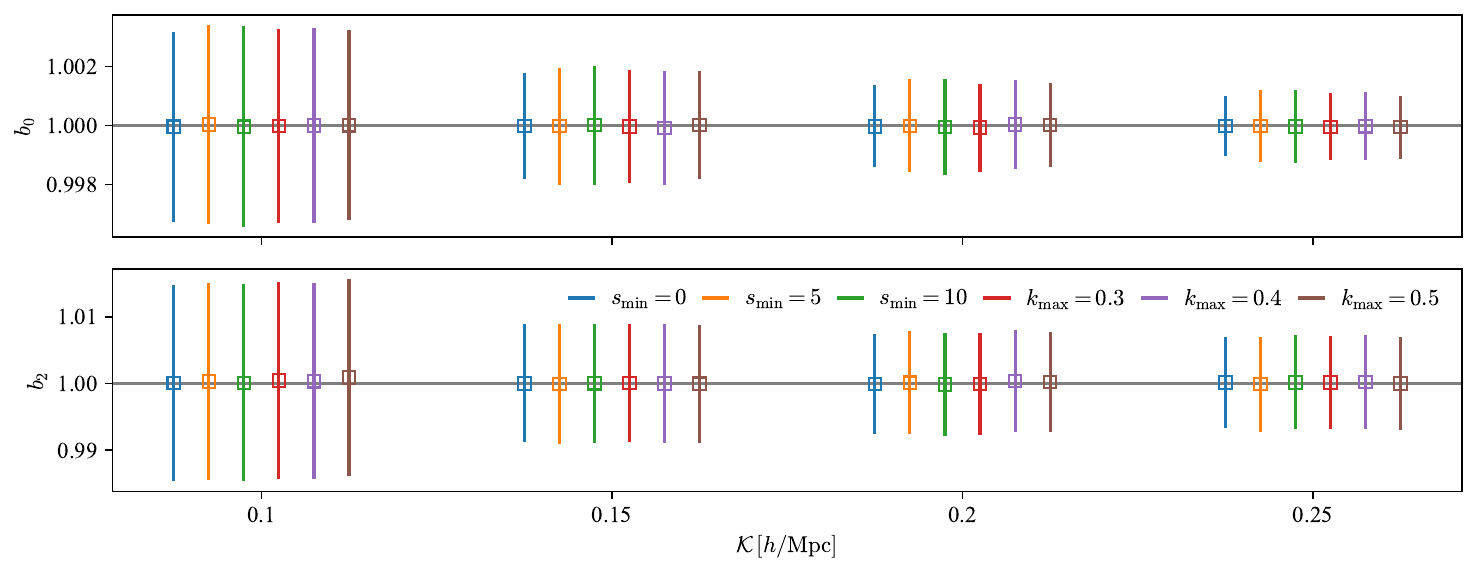}
    \caption{The average of the 123 fitting parameters ($b_\ell$) detailed in Section~\ref{sec:cov_mat_constrain_power}, obtained from 123 \textsc{SLICS} power spectra fitted on $k\in[0.02, \mathcal{K}]\,\ku$. The error bars are computed as the average of 123 $\sigma_{b_\ell}$, divided by $\sqrt{123}$, where $\sigma_{b_\ell}$ is the standard deviation of the $b_\ell$ posterior distribution. The different colours stand for the different \textsc{FastPM} covariance matrices exhibited in Figure~\ref{fig:correlation_pspec_1296_780}.} 
    \label{fig:pspec_fit_avg_b0_b2}
\end{figure*}

\begin{figure*}
	\includegraphics[width=\textwidth]{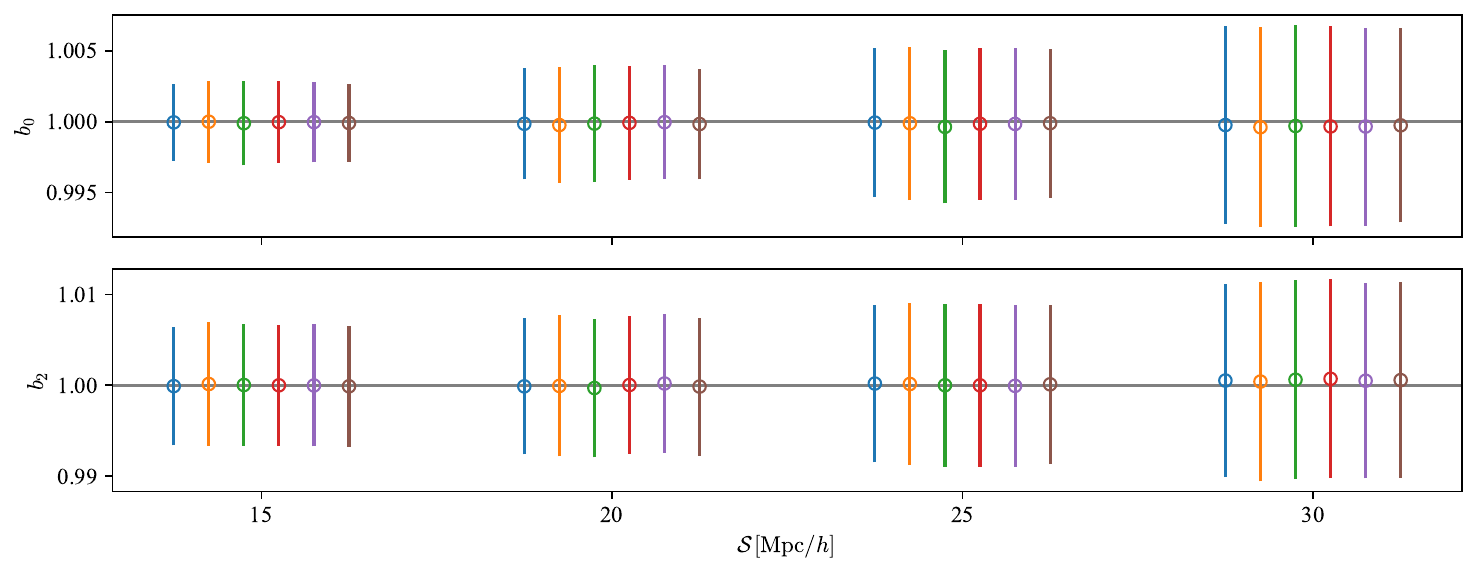}
    \caption{Same as Figure~\ref{fig:pspec_fit_avg_b0_b2}, but the fitting is performed on 123 \textsc{SLICS} 2PCF and $s\in[\mathcal{S}, 200]$ using the covariance matrices exhibited in Figure~\ref{fig:correlation_2pcf_1296_780}.}
    \label{fig:2pcf_fit_avg_b0_b2}
\end{figure*}

Figures~\ref{fig:pspec_fit_avg_b0_b2} and \ref{fig:2pcf_fit_avg_b0_b2} show the results of fitting the clustering with the simplified model detailed in Section~\ref{sec:cov_mat_constrain_power}.  Since most results are consistent with the expected value of one, we only display the values for $\mathcal{K}=0.25\,\ku$ and $\mathcal{S}=20\,\su$ in the main text.

\bsp	
\label{lastpage}
\end{document}